\documentclass[pra,reprint,superscriptaddress]{revtex4-2}

\usepackage{graphicx}
\usepackage{enumitem}
\usepackage{amsmath}
\newcommand{\ket}[1]{|{#1}\rangle}
\newcommand{\bra}[1]{\langle{#1}|}

\usepackage{amsfonts}

\usepackage[usenames,dvipsnames]{xcolor}
\usepackage[colorlinks=true,citecolor=Cerulean,linkcolor=RubineRed,urlcolor=Cerulean]{hyperref}

\usepackage{qcircuit}
\usepackage{placeins}

\begin{document}

\title{Tensor network approaches for plasma dynamics}

\author{Ryan J. J. Connor}
\email{ryan.connor.2019@uni.strath.ac.uk}
\affiliation{Department of Physics and SUPA, University of Strathclyde, Glasgow G4 0NG, United Kingdom}
\author{Preetma Soin}
\affiliation{AWE, Aldermaston, Reading, RG7 4PR, United Kingdom}
\author{Callum W. Duncan}
\affiliation{Aegiq Ltd., Cooper Buildings, Arundel Street, Sheffield, S1 2NS, United Kingdom}
\affiliation{Department of Physics and SUPA, University of Strathclyde, Glasgow G4 0NG, United Kingdom}

\author{Andrew J. Daley}
\affiliation{Department of Physics, University of Oxford, Clarendon Laboratory, OX1 3PU Oxford, UK}
\affiliation{Department of Physics and SUPA, University of Strathclyde, Glasgow G4 0NG, United Kingdom}

\date{\today}

\begin{abstract}

The dynamics of plasmas are governed by a set of non-linear differential equations which remain challenging to solve directly for large 2D and 3D problems. Here we investigate how tensor networks could be applied to plasmas described by the Vlasov-Maxwell system of equations and investigate parameter regimes which show promise for efficient simulations. We show for low-dimensional problems that the simplest form of tensor networks known as a Matrix Product State performs sufficiently well, however in regimes with a strong permanent magnetic field or high-dimensional problems one may need to consider alternative tensor network geometries. We conclude the study of the Vlasov-Maxwell system with the application of tensor networks to an industrially relevant test case and validate our results against state of the art plasma solvers based on Particle-In-Cell codes. We also extend the application of tensor networks to the alternative plasma description of Magnetohydrodynamics and outline how this can be encoded using Matrix Product States.

\end{abstract}

\maketitle
\section{Introduction}

The modelling and simulation of the complex phenomena which can occur when a plasma interacts with external and self-generated electric and magnetic fields is a central task across many disciplines. A widely used framework for the descriptions of plasmas is that of the Vlasov-Maxwell kinetic model \cite{JUNO2018110,VALENTINI2007753,Bellan_2006}, where the plasma is described by a distribution function in phase space describing the number of electrons and ions present in a region of space with a given velocity. This kinetic modelling of plasmas is frequently used to study laser-plasma interactions \cite{LI2023111733,ZHANG2024108932,Beat_paper} and turbulence in plasmas \cite{Nastac_2024}. An alternative description of plasmas is the Magnetohydrodynamic (MHD) model. The governing MHD equations more closely resemble the Euler equations for compressible fluid flows, but with additional terms to account for electric and magnetic fields \cite{MHD_notes,Bellan_2006}, and can be viewed as moments of the Vlasov equation. The MHD formalism is widely used, in particularly for stellar formation and large scale turbulence modelling \cite{Dynamo_Bott,Tran_Yu_Blackbourn_2013,li2025dataconstrained3dmhdsimulation,migdal2025dualtheorymhdturbulence,iwasaki2025comparativeanalysishalleffect}.


In recent years there has been an exploration of how alternative simulation approaches based on tensor networks could be used to simulate non-linear differential equations, in particular for fluid dynamics \cite{Q_inspired_fluids,Wall_flows,hölscher2024quantuminspiredfluidsimulation2d}, models of cold-atoms \cite{multigrid,GPE_paper} and simulations of high-dimensional distribution functions \cite{gourianov2024tensornetworksenablecalculation}.  Tensor networks were originally developed as a tool in the quantum-many body field \cite{biamonte2017tensornetworksnutshell,TN_quantum_systems,Or_s_2014} where they have been used to probe low lying energy states of interacting particles through the now famous density matrix renormalisation group algorithm (DMRG) \cite{Schollw_ck_2011,chan2016matrixproductoperatorsmatrix}, and even to study out of equilibrium phenomena via performing time-evolution of an initial quantum state under the action of a Hamiltonian  \cite{Paeckel_2019,Vanderstraeten_2019,TDVP,Schollw_ck_2011}.  Tensor networks provide a mechanism to achieve physically motivated data compression, allowing one to simulate large systems of many-interacting particles by limiting the amount of entanglement which is present in the representation \cite{Evenbly_2011,cichocki2014tensornetworksbigdata,schuch2008}. The simplest and most commonly used tensor network is that of the Matrix product state, also known as a tensor train. More complex tensor networks are available which one could exploit, however these will generally result in more numerically expensive calculations \cite{comb_tn}.  

By extending the application of tensor networks to simulations of classical high-dimensional partial differential equations, physically motivated data compression is achieved through discarding correlations which do not contribute to the overall physics, as opposed to discarding important length scales as one would do if using an overly coarse simulation grid. This in principle can allow tensor networks to simulate high-dimensional problems directly on dense grids provided one can restrict correlations building up within the representation. Additionally, tensor networks are intimately linked to quantum circuits, and thus may provide a route to the simulations of non-linear partial differential equations (PDEs) on quantum hardware \cite{Qalg_CFD,Var_algs,Tensor_prog_Qalg,Cirac_MPS_unit}.  Previous works have begun to explore the application of tensor networks to the simulations of plasma dynamics \cite{Plasma_MPS,Plasma_comb}. In these works tensor networks in the form of Matrix Product States (MPS) were used to demonstrate that one can perform efficient simulations of the Vlasov formalism with small bond dimensions for some example test cases, and an alternative comb-based tensor network was introduced. Thus far, studies into the application of tensor networks for plasma physics have focused on simulating the Vlasov equation \cite{Plasma_MPS,Plasma_comb}.

In this work we further the application of tensor networks to the modelling of the Vlasov equation, first exploring a new test case which can be represented by an MPS with small bond dimensions, and then providing a systematic analysis of the response of the bond dimension as a function of externally applied magnetic fields. We find that applying an external magnetic field to a plasma distribution leads to a rapid increase in the required bond dimension as compared to the case of no magnetic field. Inspired by Ref.~\cite{Plasma_comb} and \cite{TTN_miles_func} we also make use of the comb tensor network, and additionally provide a direct comparison between these different tensor network geometries in an industrially relevant problem setting, where we find that using a comb geometry produces results equivalent to a larger bond dimension MPS. Furthermore we demonstrate the application of tensor networks to the MHD description of plasma dynamics and show that one can perform simulations using the MPS geometry, achieving data compression whilst still capturing the physical features such as shock formation.

The outline of this paper is as follows: In Section \ref{sec:Sim_plas_TN} we outline the specifics for modelling plasmas and outline how tensor networks may be exploited, which is then applied to a demonstration test case in Section \ref{sec:Bump_on}. Section \ref{sec:mag_anals} outlines the behaviour of the required bond dimension of an MPS as a function of external magnetic field for in a chosen problem, before we consider an alternate tensor network geometry in section \ref{sec:Beat} for our industrial test case. Finally in Section \ref{sec:MHD} we present the alternative description of plasmas based on the MHD framework, showing that the simplest tensor network geometry is sufficient for these simulations.

\section{Simulating Plasmas with Tensor Networks} \label{sec:Sim_plas_TN}
The Vlasov-Maxwell equations are a set of differential equations which can be used for modelling the behaviour of plasmas \cite{JUNO2018110,VALENTINI2007753,Bellan_2006}.  These systems of equations combine Maxwell's equations for the propagation of Electromagnetic (EM) fields with the kinetic Vlasov equation which governs the evolution of a phase space distribution function $f_s(\mathbf{x},\mathbf{v},t)$ of a charged species $s$, defined as the number of particles in a unit of phase space at a given time $t$, where $\mathbf{x}$ and $\mathbf{v}$ represent the position and velocity respectively. The number density for each species is thus given by $n_s(\mathbf{x},t)=\int_{-\infty}^{\infty} f_s(\mathbf{x},\mathbf{v},t) d\mathbf{v}$. For each species within the plasma, the Vlasov equation, neglecting collisions between the particles, takes the form of \cite{Bengt},
\begin{align}
    \frac{\partial f_s(\mathbf{x},\mathbf{v},t)}{\partial t} + &\mathbf{v}\cdot \nabla_r f_s(\mathbf{x},\mathbf{v},t) \nonumber + \\ &\frac{q_s}{m_s} [\mathbf{E}+\mathbf{v}\times\mathbf{B}]\cdot \nabla_v f_s(\mathbf{x},\mathbf{v},t)=0 \ , \label{eq:Vlasov}
\end{align}
with $q_s$ each particle species charge and $m_s$ its mass. $\mathbf{E} , \mathbf{B}$ are the electric and magnetic fields respectively. The EM fields are governed by the Ampere-Maxwell and Faraday equations,
\begin{eqnarray}
    \frac{\partial \mathbf{E}}{\partial t} = c^2 \nabla \times \mathbf{B} - \frac{1}{\epsilon_o} \mathbf{J} \ , \label{eq:E_dt}\\ 
     \frac{\partial \mathbf{B}}{\partial t} =- \nabla \times \mathbf{E} \ ,\label{eq:B_dt} 
\end{eqnarray}
respectively. The current density $\mathbf{J}$ is found via,
\begin{equation}
    \mathbf{J} = \sum_s q_s \iint_{\mathbf{v}} \mathbf{v} f_s(\mathbf{x},\mathbf{v},t) d\mathbf{v} \ ,
\end{equation}
where one sums over all charged particle species $s$ present to obtain the net current density within the plasma. Throughout this paper we work with non-dimensional units by setting the speed of light $c$, permittivity of free space $\epsilon_o$, mass of electron and charge of the proton all to unity. 

The divergence-less condition for the magnetic field and Gauss' law for the electric field do not appear explicitly, but it can be shown that they are fulfilled at all future times if satisfied initially. 


This full phase space model naturally presents a challenge to conventional simulations of plasmas, where for a simulation of a three dimensional plasma one must work with $6+1$ dimensions, three spatial dimensions, three velocity dimensions and the time dimension. Directly simulating these systems with a fine enough resolution to resolve dynamics is a daunting and numerically intensive task, and thus is rarely carried out. Instead one often reverts to alternative simulation approaches such as Particle-in-cell (PIC) codes \cite{PIC_overview,banerjee2023particleincellcodecomparisonion}. PIC simulations are a statistical based approach where one models the plasma as being comprised of a collection of computational particles which move on continuous paths in phase space, whilst the EM fields are defined on a fixed grid \cite{GERMASCHEWSKI2016305}. The resolution of PIC methods controlled by the number of computational particles, with more computational particles per cell leading to an increased resolution but at a larger computational cost. As this is a statistical method, there are numerical fluctuations in the results which scale as the inverse of the square root of the number of computational particles, and thus in some cases one may need a large number of computational particles in order to achieve a sufficiently low numerical noise rate \cite{PhysRevSTAB.18.114201,Arber_2015}.

\subsection{Encoding into MPS}
To exploit tensor networks to simulate plasma dynamics via the Vlasov-Maxwell model, our distributions functions $f_s$ and EM fields $\mathbf{E}, \mathbf{B}$ must be encoded into an appropriate tensor network geometry, the simplest of which being a Matrix product state (MPS). Encoding a function over a high-dimensional grid into an MPS has been exploited in previous studies of fluid and plasma dynamics with tensor networks \cite{Plasma_comb,Plasma_MPS,Q_inspired_fluids,gourianov2024tensornetworksenablecalculation}, and we outline only the basics here. Each dimension in the problem at hand is discretised into a grid of $2^{N_l}$ grid points labelled $q_i^l$, with $i\in [0,1,2,\cdots,2^{N_l}-1]$ where $l$ labels the dimension, $l\in[x,y,v_x,v_y \cdots]$. Note that one does not require a uniform number of grid points across each dimension, and infact most often we assume that each dimension is discretised into a different number of grid points.  We then break down the grid-point index $q_i^l$ into its binary representation,
\begin{equation}
    q_i^l = \left( \sigma_1^l, \sigma_2^l, \dots, \sigma_{N_l}^l \right),
\end{equation}
where each $\sigma_j^l$ can be either $0$ or $1$. These binary coefficients can be viewed as length scales in each dimension. $\sigma_1^l$ informs us if a grid point is in the first half of the grid or the second half, $\sigma_2^l$ informs of which quarter of the grid a grid-point is located in, and so on, with $\sigma_{N_l}^l$ representing the finest scale.

Rewriting the particle distribution functions in this way presents it as an $Nth$ order tensor, where  $N=\sum_l N_l$,
\begin{equation}
    f_s(x,y,v_x,v_y,\cdots) = {f_s}_{ \sigma_1^x, \dots, \sigma_{N_x}^x,\sigma_1^y, \dots, \sigma_{N_y}^y, \cdots } \ .
\end{equation}
By applying singular value decompositions (SVDs) onto the tensor ${f_s}_{ \sigma_1^x, \dots, \sigma_{N_x}^x,\sigma_1^y, \dots, \sigma_{N_y}^y, \cdots } $, and restricting the number of singular values retained, one arrives at the MPS representation,
\begin{equation}
    f_s = \sum_{\alpha_1}^{\chi_1} \sum_{\alpha_2}^{\chi_2} \dots \sum_{\alpha_{N-1}}^{\chi_{N-1}} M_{\alpha_1}^{\Tilde{\sigma}_1} M_{\alpha_1,\alpha_2}^{\Tilde{\sigma}_2} \dots M_{\alpha_{N-1}}^{\Tilde{\sigma}_{N}} \ , \label{eq:MPStensor}
\end{equation}
where $M^{\tilde{\sigma}_n}_{\alpha_{n-1},\alpha_n}$ denotes a local third order tensor with two bond indices $\alpha_{n-1}$ and $\alpha_n$, and the physical binary index $\tilde{\sigma}_n$. The $\alpha_n$ indices are known as bonds, and the number of these parameters $\chi_n$ is known as the bond dimension. These are responsible for carrying correlations within the MPS, and in general the larger $\chi_n$ at any given bond the greater the amount of correlations between the two halves of the MPS at that bond. We have additionally introduced the following notation to simplify Eq.~\eqref{eq:MPStensor}
\begin{align}
 \Tilde{\sigma}_n=\left\{\begin{array}{ll}
                  \sigma_n^x, & 1\le n\le N_x,\\[0.1cm]
                  \sigma_{n-Nx}^y, & N_x< n \le N_x+N_y,\\[0.1cm]
                   \sigma_{n-N_x-N_y}^{v_x}, & N_x+N_y< n \le N_x+N_y+N_{v_x} \ . 
                 \end{array}\right.
\end{align}
This results in an MPS of total length $N$, where each tensor represents a length scale within a dimension. Likewise all required operations such as spatial derivatives can be encoded in the operator equivalents known as Matrix Product Operators (MPOs) \cite{Q_inspired_fluids,wall_bounded_flows,Plasma_MPS,Plasma_comb}.

\begin{figure}[htb!]
    \centering
    \includegraphics[width=.9\linewidth]{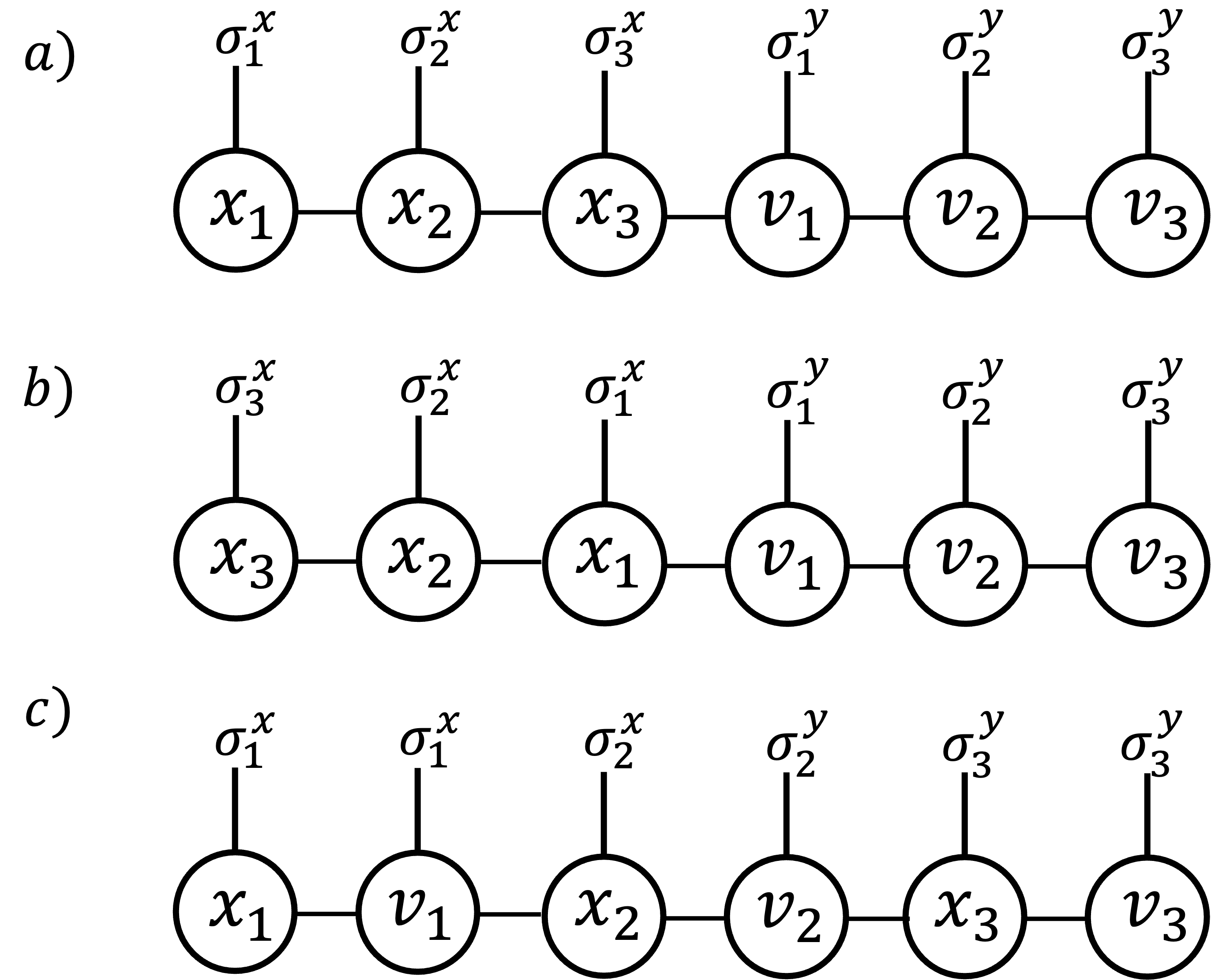}
    \caption{Three possible MPS decomposition for a 1D1V plasma system. We show two possible sequential orderings in $a)$, where the $x$ dimension is encoded from coarsest to finest scale followed by the velocity dimension in same order, and $b)$, where the $x$ dimension is encoded from finest to coarsest and the velocity from coarsest to finest. $c)$ shows an inter-leafed ordering where one alternates between an $x$ and $v$ tensor, but grouping together length scales of each dimension. }
    \label{fig:Example_MPS_encodings}
\end{figure}

There are many possible ways to decompose a high-dimensional function such as $f_s$ into an MPS, with the above a single example. In the above encoding each dimension is encoded in turn, with the tensors within each dimension always ordered from the coarsest scale $\sigma_1^l$ to the finest scale $\sigma_{N_l}^l$. This need not be the case, for example one can arbitrarily change the ordering of the length scales, or could alternate tensors between dimensions. A few alternate example encodings for a 2D phase space ($x$ and $v$) are demonstrated pictorially in Fig.~\ref{fig:Example_MPS_encodings}. The optimal encoding one can use is that which reduces the bond dimension required to accurately represent the functions in MPS form, and this arises for the ordering with the minimal correlations across a bi-partition of the MPS at each bond.

\section{Bump on tail instability} \label{sec:Bump_on}

\begin{figure*}[t]
    \centering
    \includegraphics[width=\linewidth]{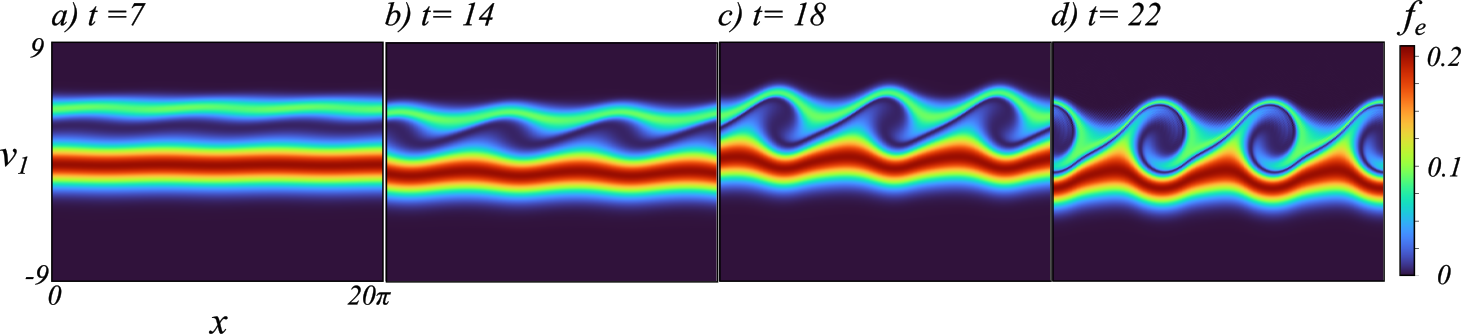}
    \caption{Snapshots of a cut of the electron distribution $f_e(x,v_1)$ through phase space for the bump-on-tail instability. We show a slice through the $v_2=0$ plane, resulting in a 2D function in $x$ and $v_1$, at various simulation times of  $a) t=7$, $b) t=14$, $c) t=18$ and $d) t=22$. These simulations where conducted on a $2^8 \times 2^8 \times 2^8$ phase space grid, resulting in a $24$ site MPS encoding, and evolved via a fixed time-step $4th$ order Runge kutta scheme with a time-step of $dt=0.0014$. A truncation cutoff of $10^{-15}$ was used to adaptively control the bond dimension $\chi$. }
    \label{fig:Bump_snaps}
\end{figure*}

We demonstrate the ability of an MPS to simulate plasma physics by considering the case of a 1D2V Bump-on-tail plasma instability \cite{Bump_on_tail} where we have one spatial dimension, $x$, and two velocity dimensions, $v_1$ and $v_2$. One can think of this scenario physically as having a completely uniform distribution in the $y$ direction, and thus neglecting the $y$ dependence. We consider mobile electrons with $q_e=-1$ and $m_e=1$ and with the immobile ions forming a neutralizing background. The dynamics of this test case evolve according to, 
\begin{equation}
    \frac{\partial f_e(x,v_1,v_2,t)}{\partial t} + v_1 \frac{\partial f_e}{\partial x} + \frac{q}{m}(\mathbf{E} + \mathbf{v}\times\mathbf{B}) \cdot \nabla_{\mathbf{v}} f_e = 0 \ .
\end{equation}
In this 1D2V model we take the electric and magnetic fields to have the form of,
\begin{equation}
    \mathbf{E} = \begin{pmatrix}
         E_1(x,t) \\
         E_2(x,t)\\
         0
    \end{pmatrix}  \ , \   \mathbf{B} = \begin{pmatrix}
         0 \\
         0\\
         B_3(x,t)
    \end{pmatrix} \ .
\end{equation}
The Bump-on-tail instability initially consists of a main band of electrons uniformly distributed along $x$ with a Gaussian distribution in velocity space, along with a secondary band of electrons initialised with a non-zero drift velocity along the $v_1$ dimension. Specifically we assume the following initial conditions for the electron distribution function \cite{Bump_on_tail},
\begin{widetext}
\begin{equation}
    f_e(x,v_1,v_2,t=0) = \frac{1}{\sqrt{2} \pi}(\alpha e^{\frac{-v_1^2}{2}} + \beta e^{-2(v_1-4.5)^2}(1+\gamma \cos(kx)) )e^{-v_2^2}\ ,
\end{equation}
\end{widetext}
with the parameters $\alpha=9/10, \beta=2/10, \gamma=0.03, k=0.3$ with $x=[0,20\pi]$, and $v_1=v_2=[-9,9]$. We assume periodic boundaries along the $x$ dimension and open boundaries along the velocity dimensions. The electric field is initialised at $t=0$ via solving the Poisson equation,
\begin{equation}
    \nabla \cdot \mathbf{E}= - \nabla^2 \phi = \rho \ , \label{eq:poisson}
\end{equation}
where $\rho=q_e\iint f_e dv_1 dv_2$ and $\phi$ is the electrostatic potential. We additionally initialise with $\mathbf{B}=0$. This problem is simulated on an $2^8\times 2^8 \times 2^8$ phase space grid which is sequentially encoded into a $24$ site MPS. The first $8$ tensors encode the $x$ dimension, the next $8$ tensors the $v_1$ dimension and the final tensors encode the $v_2$ dimension. Each dimension is ordered with the first tensor representing the coarsest scale and final the smallest length scale, exactly as outlined in Eq.~\eqref{eq:MPStensor}.  After setting up the problem we employ the standard $4th$ order Runge-kutta (RK4) scheme \cite{Numerical_analysis} to evolve the Vlasov equation in time. The spatial and velocity derivates are calculated via central finite difference approximations. We utilise an adaptive bond dimension during our simulations which is controlled by truncation cutoff scheme. To do this truncation one performs an SVD on each bond within the MPS and retains only the largest singular values, such that the sum of the squared singular values discarded does not exceed this cutoff. After each MPO-MPS contraction, MPS addition, and after each time-step we truncate the MPS using a cutoff of $10^{-15}$. We show snapshots of the dynamics in a cut through the $v_2=0$ plane in phase space for different simulation times in Fig.~\ref{fig:Bump_snaps}, where vortices develop in phase space between the two bands of electrons introduced.


\begin{figure}[htb!]
\centering
\includegraphics[width=\linewidth]{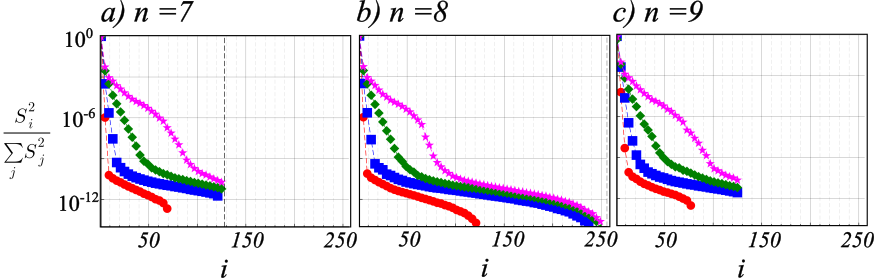}
\caption{Squared singular values of the MPS representation of the electron distribution function $f_e$ in descending order during the bump-on-tail simulation. We plot the singular values across the a)$n=7$ b)$n=8$ and c) $n=9$ bond of the MPS at simulation times of $t=7$ (red dots), $t=14$ (blue square), $t=18$ (green diamond) and $t=22$ (pink stars). At all times and at all bonds the singular values decay fast indicating the ability to effectively truncate the MPS representation. }
    \label{fig:Bump_schmidts_S1}
\end{figure}

To analyse how well a simulation is suited for treatment with tensor networks, one can analyse the singular values (also known as Schmidt coefficients) at every bond in the MPS. These singular values are the key in being able to truncate a tensor network, and thus studying the behaviour of these coefficients can inform us about the representibility of the distribution functions in a given MPS encoding. We display these singular values for the simulation presented in Fig.~\ref{fig:Bump_snaps} for different bonds $n$ in the MPS in Fig.~\ref{fig:Bump_schmidts_S1} where we plot the singular values in decaying order. Across all bonds one can clearly observe the singular values growing over time, which physically represents the generation of correlations in the data as the simulation progresses in time. The $n=8$ bond in the MPS corresponds to the bipartition between the spatial and velocity dimensions, and thus describes correlations between $x$ and $v_1,v_2$. We observe that the singular values at all times appear to decay fast in an exponential manner. This indicates that if one was to truncate the MPS to a small bond dimension $\chi$, only very small singular values would be discarded and hence one would not expect to observe significant truncation errors. 
\vspace{-2cm}
\section{Correlation Analysis} \label{sec:mag_anals}
The previous section and the works of \cite{Plasma_MPS,Plasma_comb} have demonstrated that MPS can perform well in simulations of plasmas through the Vlasov equation. It is also of equal importance to analyse how modifying various control parameters of the plasma simulation impacts the bond dimension required to accurately capture the changing physics. To this end we next consider a likewise 1D2V test case, this time probing the impact of applying external magnetic fields to the plasma. We consider again only electrons with the following initial distribution function \cite{Bengt},
\begin{equation}
    f(x,v_1,v_2,t=0)=[1+A\sin(kx)] \frac{e^{-\frac{1}{2}(v_1^2+v_2^2)}}{\sqrt{2 \pi}} \ , \label{eq:mag_init}
\end{equation}
where $A=0.001$ is the amplitude and $k=0.4$ is the wave number of a perturbation along the spatial dimension. A constant external magnetic field is directed along the $z$ axis, and the electric fields are calculated in a self consistent manner via the Poisson equation (Eq.~\eqref{eq:poisson}).

This problem was simulated for a range of external magnetic field strengths $B_{o}$ on a $2^8\times 2^8\times2^8$ phase space grid, resulting in a $24$ site MPS, with an adaptive bond dimension controlled via a relative truncation cutoff of $10^{-15}$. A fixed time-step RK4 scheme is used to perform these simulations. To perform an RK4 time evolution scheme, one must evaluate the derivative of our function, $ \frac{\partial f(t)}{\partial t}$, at four intermediate times from $t$ to $t+dt$ as follows, 
\begin{eqnarray}
    K_1 &=& \frac{\partial f(t)}{\partial t}  , \\ 
    f_1 &=& f(t) +\frac{dt}{2} K_1 , \\
    K_2 &=& \frac{\partial f(t+\frac{dt}{2})}{\partial t} , \\
    f_2 &=& f(t) +\frac{dt}{2} K_2 , \\
    K_3 &=& \frac{\partial f_2(t+\frac{dt}{2})}{\partial t} , \\
    f_3 &=& f(t) + dt K_3 ,\\
     K_4 &=& \frac{\partial f_3(t+dt)}{\partial t} .
\end{eqnarray}
One then obtains the distribution at the next time step via a weighted average of the $K_i$ intermediate derivatives,
\begin{equation}
    f(t+dt)=f(t) +\frac{dt}{6}(K_1+2K_2+2K_3+K_4) \ .
\end{equation}

\begin{figure}[t]
    \centering
    \includegraphics[width=\linewidth]{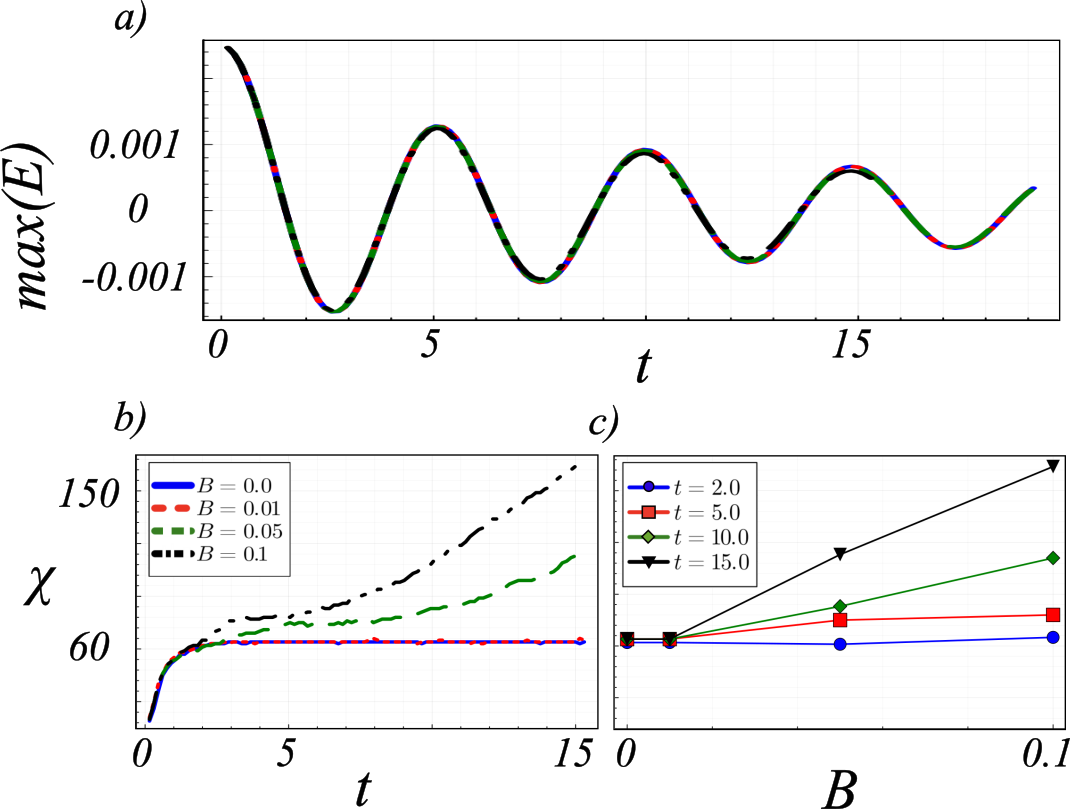}
    \caption{Results from simulation of Eq.~\eqref{eq:mag_init} on an $2^8\times2^8\times2^8$ phase space grid for various magnetic field strengths $B_o$. $a)$ Maximum of electric field and $b)$ maximal bond dimension of $K_1$ intermediate MPS as a function of simulation time $t$ for external magnetic field strengths of $B=0$ (solid blue), $B=0.01$ (dotted red), $B=0.05$ (dashed green) and $B=0.1$ (dash dotted black). $c)$ Maximal bond dimensions of $K_1$ MPS as a function of external magnetic field strength $B_o$ at simulation times of $t=3.5$ (blue dots), $t=6.1$ (red squares), $t=9.3$ (green diamonds) and $t=12.2$ (black triangles).}
    \label{fig:mag_dynamics}
\end{figure}

For each value of the external magnetic field we encode identically into a sequential MPS and time-evolve under the RK4 scheme. The resultant dynamics for each $B_o$ results in an oscillation of the maximum of the electric field which is demonstrated in Fig.~\ref{fig:mag_dynamics} $a$. Along with tracking the dynamics for each $B_o$, we additionally analyse how the required bond dimension scales with time. In Fig.~\ref{fig:mag_dynamics} $b$ the bond dimensions of $K_1$ during the simulations are plotted for each $B_o$ considered. Across all magnetic fields, we observe an initial increase in bond dimension from early times, which indicates a growth of correlations within the MPS as one would expect. However for the first case of $B_o=0$, the bond dimension eventually plateaus at a modest $\chi\sim 60$, and stops growing in time, indicating a saturation of the correlations. In contrast the non-zero magnetic fields with $B_o>0.01$ display bond dimensions which grow larger in time, without a clear saturation over the timescales considered in the present simulations, reaching $\chi=110$ for $B_o=0.05$ and $\chi=150$ for $B_o=0.1$ at $t=15$. As we increase the strength of $B_o$, the bond dimension required to represent the dynamics grows, indicating an increase in the correlations present within the MPS. This can be further observed in Fig.~\ref{fig:mag_dynamics} $c$ where we plot the maximal bond dimensions at fixed simulation times as a function of magnetic field strength $B_o$. At all simulation times, increasing $B_o$ results in a larger bond dimension, and at late simulation times of $t\ge10$ a linear growth in $B_o$ produces a linear increase in the required bond dimension. Furthermore, the rate of increase in $\chi$ with $B_o$ appears to grow with simulation time, i.e. at later times there is a larger impact on the bond dimension depending on the magnetic field strength. 

\begin{figure}[t]
    \centering
    \includegraphics[width=0.8\linewidth]{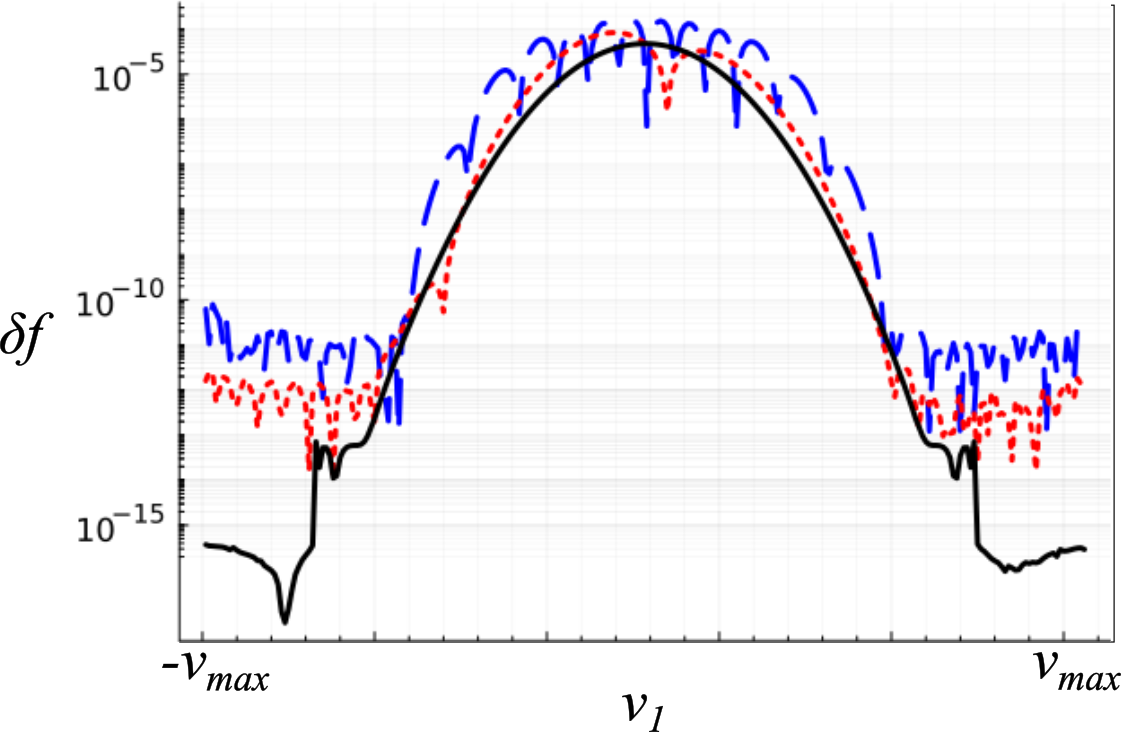}
    \caption{Results from simulation of Eq.~\eqref{eq:mag_init} on an $2^8\times2^8\times2^8$ phase space grid for various magnetic field strengths $B_o$. We plot a slice through phase space along the $x=\frac{L}{2}$, $v_2=0$ plane of the change in the electron distribution relative to the initial condition $\delta f= f(t)-f(t=0)$ at a simulation time of $t=15$. We show this slice for the case of $B_0=0$ (solid black), $B_o=0.01$ (dotted red) and $B_o=0.1$ (dashed blue).}
    \label{fig:mag_slice}
\end{figure}

For each of these different magnetic fields one can plot a slice through phase space at a fixed time to compare how the magnetic field impacts the electron distribution function. In Fig.~\ref{fig:mag_slice} we show such slices along the $x=\frac{L}{2}$, $v_2=0$ plane. The zero magnetic field case produces a simple Gaussian-like profile along this plane, reflecting the limited change from the initial condition, which thus requires only small bond dimensions to represent. In contrast for the case of $B=0.01$ and particularly $B=0.1$, we observe a lot more structure in the distribution functions, with an overall Gaussian-like profile but permuted with apparent oscillations on-top. Additionally we observe a small amount of noise present in the non-zero magnetic field simulations at the $10^{-11}$ level which is not present in the zero magnetic field case. We expect that by improving accuracy of the simulations by using a combination of a finer grids, smaller time-steps and smaller truncation errors, we would be able to reduce this noise back to machine precision levels.

\begin{figure}[htb!]
\centering
\includegraphics[width=\linewidth]{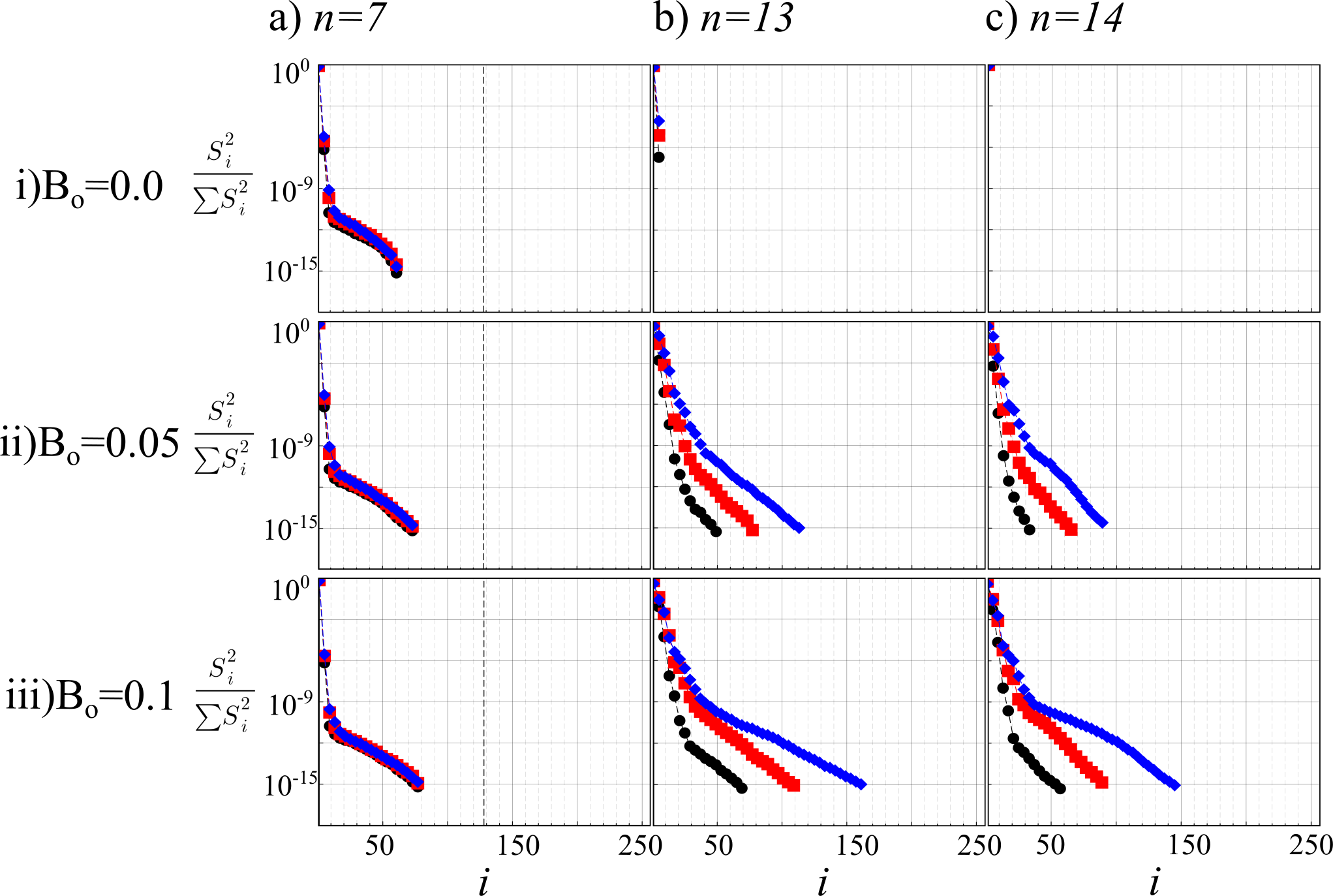}
\caption{Squared singular values of the MPS representation of the MPS $K_1$ in descending order during the plasma simulations of Eq.~\eqref{eq:mag_init}. We plot the singular values across the a)$n=7$ b)$n=13$ and c) $n=14$ bond of the MPS at simulation times of $t=5$ (black dots), $t=10$ (red square), and $t=15$ (blue diamonds), for magnetic field strengths of $i)B_o=0$, $ii) B_o=0.05$ and $B_o=0.1$. The number of present singular values exactly determines the bond dimension at the given bond, and thus rapidly decaying singular values indicates the ability to effectively truncate the MPS representation. }
    \label{fig:Mag_schmidts}
\end{figure}

We can additionally observe the impact of the magnetic field via analysing the singular values present within the various bonds of the MPS representing $K_1$ at different times, as shown in Fig.~\ref{fig:Mag_schmidts}. One can see that as time progress and as one applies a stronger magnetic field to the plasma, each singular value increases and the rate of decay reduces, again indicating a strong growth of correlations which reflects the rapidly growing bond dimension to represent the plasma distribution functions in the presence of increasingly strong external magnetic fields. This means that, for these given encodings into an MPS, larger external magnetic fields would thus require greater computational resources to represent to the same desired accuracy level within the tensor network format, as the bond dimension cannot be truncated to a small value to achieve data compression without removing non-negligible singular values and thus severely impacting the accuracy of the simulation.

\begin{figure}[t]
    \centering
    \includegraphics[width=\linewidth]{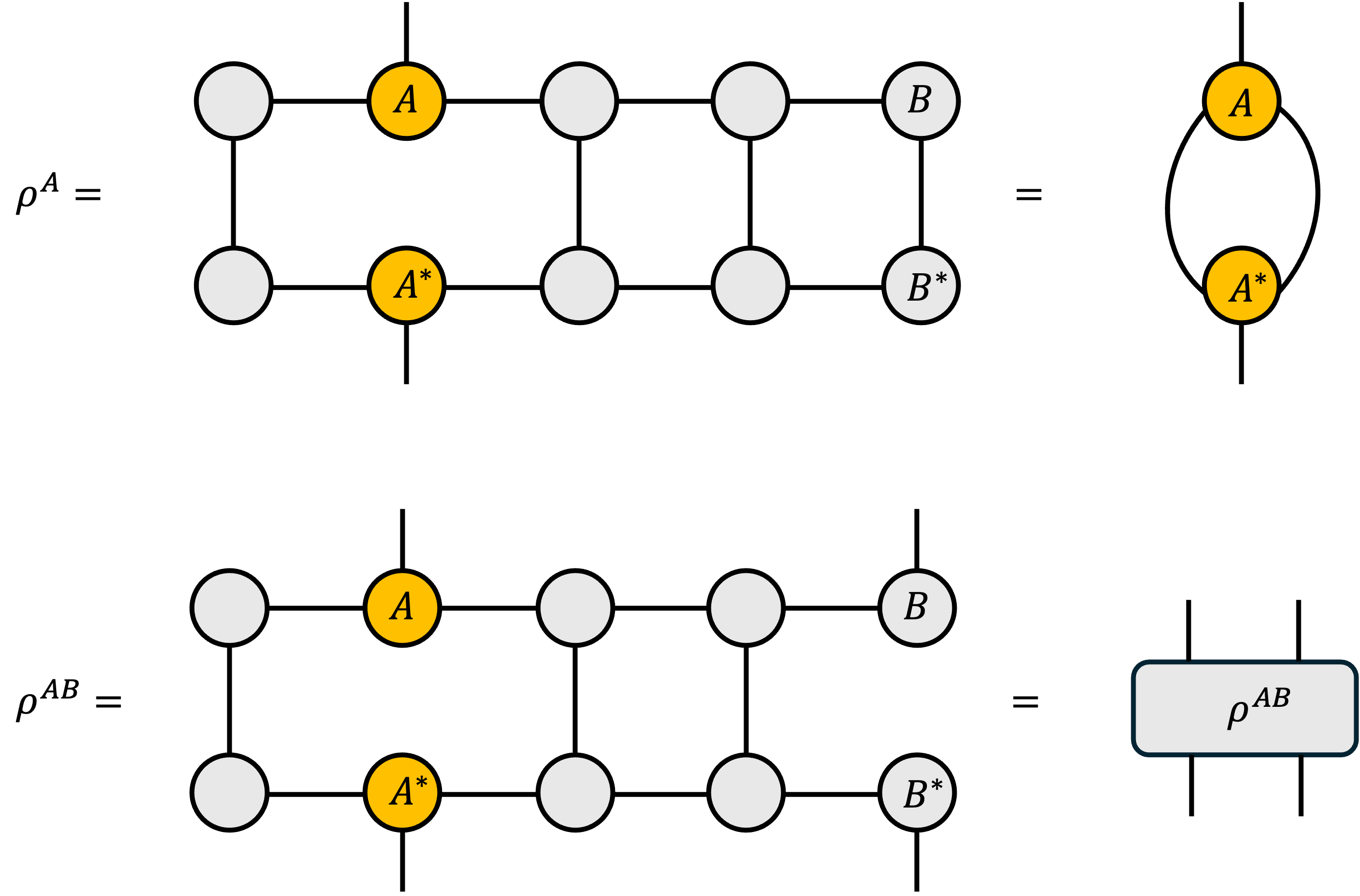}
    \caption{Tensor network diagrams for forming the required reduced density matrices to calculate the mutual information between two tensors $A$ and $B$. To simplify the calculation, one should orthogonalise the MPS around the $A$ or $B$ tensors, where gold Tensors shows the location of the orthogonality centre of the MPS.}
    \label{fig:MI_TN}
\end{figure}

\begin{figure*}[t]
    \centering
    \includegraphics[width=\linewidth]{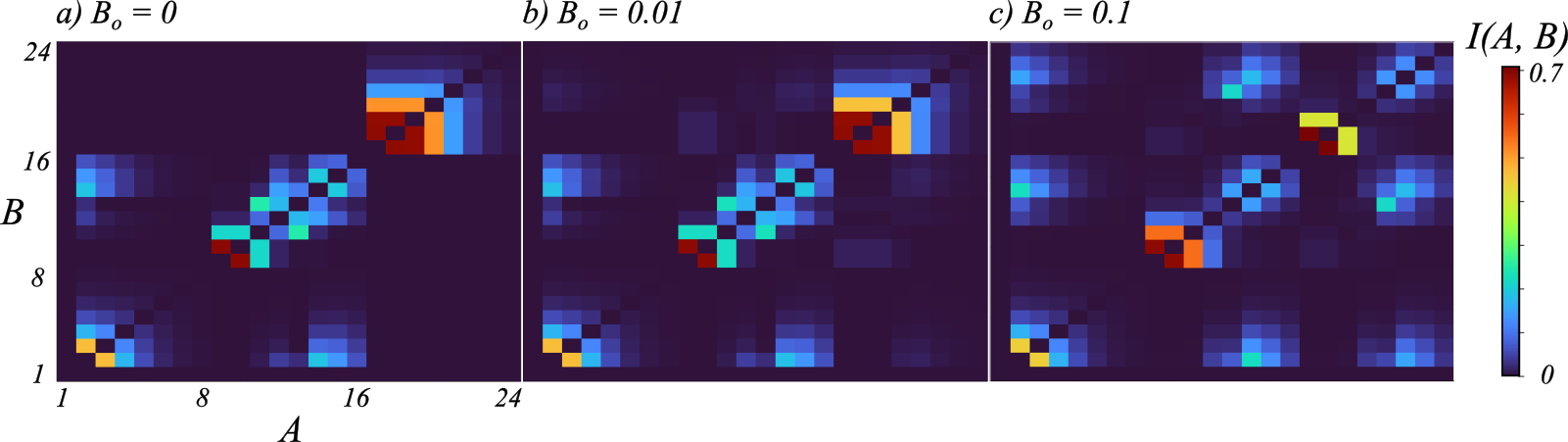}
    \caption{Plots of mutual information $I(A,B)$ between each combination of tensors $A$ and $B$ in the MPS. Note that $I(A,B)=I(B,A)$, leading to symmetric plots. We show the mutual information for the cases of $a)B_o=0$, $b) B_o=0.01$ and $c) B_o=0.1$ at a fixed simulation time of $t=15$. In the demonstrated encoding, tensors $1-8$ encode the $x$ dimension, tensors $9-16$ encode the $v_1$ dimension and $17-24$ encode the $v_2$ dimension.}
    \label{fig:MI_mag}
\end{figure*}

Analysis of the singular values does allow one to easily tell how suitable a given MPS encoding is to truncation, however it does not directly inform us which tensors within the MPS are strongly correlated to one another, instead providing information about the correlations at any given bipartition of the MPS. To this end we introduce the concept of the mutual information $I(A,B)$ as a way to gain further insight into how correlations develop within the MPS. The mutual information between two tensors $A$ and $B$ is defined as,
   \begin{equation}
        I(A,B)= S(\rho^A)+S(\rho^B) -S(\rho^{AB}) \ , 
    \end{equation}
where $S(\rho^A)$ denotes the entanglement entropy of the reduced density matrix of only tensor $A$, and $S(\rho^{AB})$ is that of the reduced system of tensors $A$ and $B$. These reduced density matrices are illustrated in diagrammatic notation in Fig.~\ref{fig:MI_TN}. One calculates the entanglement entropy of a density matrix using the Von-Neuman entropy defined as, 
\begin{equation}
    S(\rho)= -Tr \ (\rho \log{\rho} ) \ .
\end{equation}
We calculate the mutual information between each pair of tensors for the various magnetic field strengths and display the resultant distributions in Fig.~\ref{fig:MI_mag}. In the $B=0$ case, the mutual information is highly local in the sense that only neighbouring tensors appear to show strong correlations. The only substantial far reaching correlations are those between the $x$ and $v_1$ dimensions. Note that the $v_2$ dimension appears effectively decoupled in the $B=0$ case, not showing any correlations to the $x$ or $v_1$ dimensions. In contrast for the non-zero magnetic fields we observe non-zero mutual information between far separated tensors. There is now a clear coupling between the $x$ and both $v_1$ and $v_2$ dimensions, and strong correlations between $v_1$ and $v_2$ themselves. These plots are further illustrations that stronger external magnetic fields in this example seem to be creating correlations between many different regions within the MPS encoding, requiring more computational resources to accurately represent. 

\section{Beat wave}\label{sec:Beat}

\begin{figure*}[htb!]
    \centering
    \includegraphics[width=\linewidth]{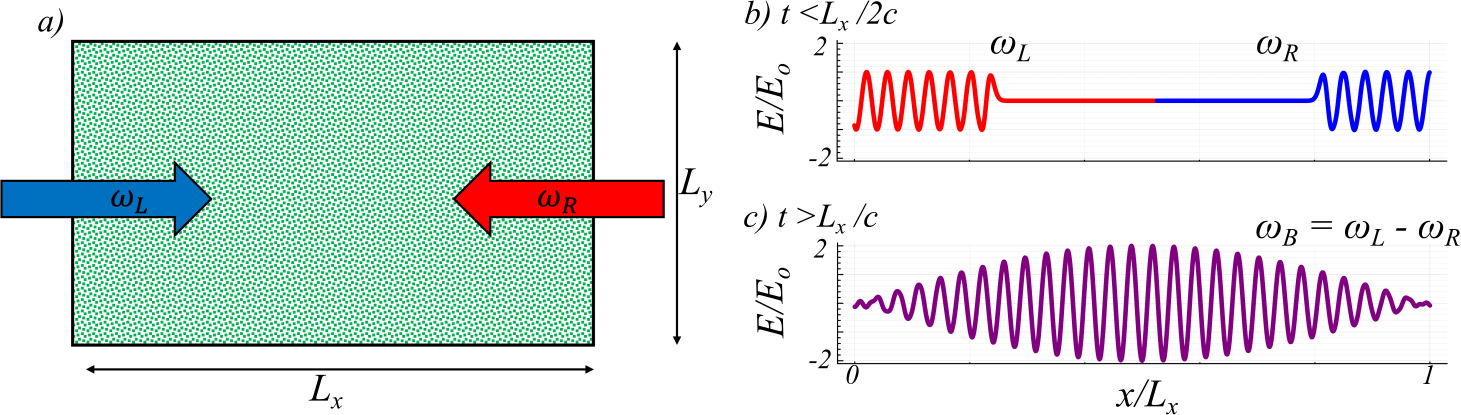}
    \caption{Illustration of beat wave problem. Two opposing lasers of frequencies $\omega_L$ and $\omega_R$ are shone onto a uniform plasma held in a rectangular box of size $L_x\times L_y$ as depicted in $\mathrm{a})$. The electric field patterns are shown in $\mathrm{b})$ and $\mathrm{c})$ for early and late times respectively. After the two counter-propagating waves overlap at a simulation time of $t=L_x/2c$, a  a beat pulse develops, where the envelope of the beat wave travels in the system with a frequency $\omega_B=\omega_L-\omega_R$. }
    \label{fig:Beat_cartoon}
\end{figure*}

As a final demonstration of the application of tensor networks to the Vlasov equation, we consider an industrially relevant test case known as a beat wave problem, shown in Fig.~\ref{fig:Beat_cartoon}. This problem consists of a rectangular slab, of dimensions $L_x \times L_y$ filled uniformly with electrons and ions. Two counter-propagating laser beams are then directly shone onto the plasma, with a frequency $\omega_L$ from the left boundary, and a frequency of $\omega_R$ from the right hand side boundary. Both lasers have the same Electric field amplitude $E_o$, however with different frequencies with $\omega_L>\omega_R$. 

As the lasers interfere, a periodic modulation of the intensity ensues with local beat pulses travelling with an envelope frequency of $\omega_{B}=\omega_L-\omega_R$ \cite{Beat_paper}, illustrated in Fig.~\ref{fig:Beat_cartoon}. An identical phenomenon occurs in the field of acoustics where simultaneously playing two slightly different frequency tones results in one hearing a beating pattern.


We define the initial distribution functions of both electrons and positive ions to be uniform in space, and have a standard Maxwellian distribution for the velocities of both the electron and ionic species,
\begin{equation}
    f_s(x,y,v_x,v_y,t=0)= \frac{1}{2 \pi v_{th_s}^2} e^{-(\mathbf{v_s})^2/(2v_{th_s}^2)}  \ ,
\end{equation}
where $v_{th_s}=\sqrt{\frac{k_B T}{m_s}}$ defines the thermal velocity for the plasma species $s$ at temperature $T$. We take the mass of the positive ion as $m_p=7300m_e$, and the plasma temperature as $k_bT=0.4 m_e c^2$ where $k_b$ is the Boltzmann constant. We have set the speed of light $c$, permittivity of free space $\epsilon_0$ and the mass of an electron $m_e=1$ to unity, and defined the laser amplitude to be $E_o=1$. This setup will correspond physically to a plasma of density $2.8\times 10^{16}\mathrm{cm^{-3}}$ with a laser intensity of $3.5\times10^{12}\mathrm{Wcm^{-2}}$ .We should note that the above parameters will lead to fast moving electrons close to the speed of light, however we leave a fully relativistic treatment with tensor networks to future work. As the lasers propagate through the plasma they cause the electrons and ions to begin to oscillate away from the original distribution function. At any later time we can write the distribution functions $f_s(x,y,v_x,v_y,t)$ in terms of the original functions at $t=0$ and a time-dependant fluctuation away from these initial configurations, $\tilde{f}_s(x,y,v_x,v_y,t)$,
\begin{equation}
   f_s(x,y,v_x,v_y,t) =  f_o(x,y,v_x,v_y) + \tilde{f}_s(x,y,v_x,v_y,t) \ .
\end{equation}
This allows one to re-express the Vlasov equation of Eq.~\eqref{eq:Vlasov} in terms of the fluctuation term,
\begin{widetext}

\begin{equation}
    \frac{\partial \tilde{f}_s(x,v,t)}{\partial t } + \mathbf{v}\cdot\nabla \tilde{f}_s(x,v,t) +\mathbf{F} \cdot \nabla_v \tilde{f}_s(x,v,t) = -\left[  \mathbf{v}\cdot \nabla f_o(x,v) + \mathbf{F}(t)\cdot \nabla_v f_o(x,v) \right] \ , \label{eq:fluc}
\end{equation}

\end{widetext}
where $\mathbf{F}(t)=\frac{q}{m}(\mathbf{E}+\mathbf{v}\times \mathbf{B})$ is the standard Lorentz force term in the Vlasov equation. The left hand side of Eq.~\eqref{eq:fluc} is exactly the standard Vlasov equation, however one has now introduced a non-zero right hand side which accounts for the effects of $f_o$. The $\mathbf{v}\cdot \nabla f_o(x,v)$ term is simply a constant term at each time-step, whereas the $\mathbf{F}(t)\cdot \nabla_v f_o(x,v)$ term must be recalculated each step. 

\subsection{Tensor network encoding} \label{sec:Beat_TN}

We consider using two tensor networks for the simulation of the beat wave problem, firstly the MPS encoding as previously prescribed. Secondly we take inspiration from the work of Ref \cite{Plasma_comb} and examine if any benefit can be obtained from moving to a more complex geometry in the form of a comb tensor network. Although using alternative tensor network geometries may increase the computational complexity, one may find that they are better suited to representing the different correlations structures which develop during complex plasma simulations as compared to MPS. Our simulations are conducted on an $2^9\times2^7\times2^8\times2^8$ phase space grid, which will result into encoding into a tensor network consisting of 32 physical indices.

\begin{figure}[htb!]
    \centering
    \includegraphics[width=\linewidth]{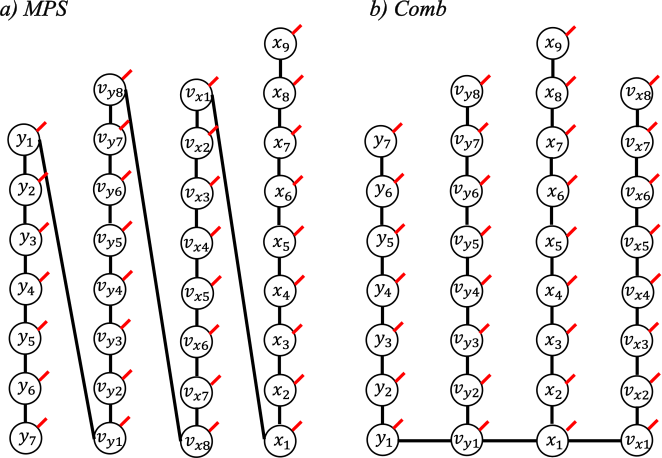}
    \caption{Illustration of our choice of possible encodings of $f(x,y,v_x,v_y)$ on an $2^9\times 2^7\times2^8\times2^8$ phase space grid into an $a)$ MPS and $b)$ comb tensor network geometry. We choose to label each tensor according to which dimension it belongs to, and the corresponding length scale of  that dimension. E.g. $x_1$ encodes the coarsest length scale of the $x$ dimension, and $v_{y_8}$ encodes the finest length scale of the $v_y$ dimension. Bond indices are shown as black lines connecting two tensors, whilst the physical indices are shown in red emerging diagonally upwards from each tensor. }
    \label{fig:Beat_encodings}
\end{figure}

Various orderings of the tensors for the MPS simulations were tested before we settled on the following: the first 7 tensors in the chain encode the $y$ dimension, starting from the finest and ending in the coarsest scale, the following 8 tensors encode the $v_y$ dimension, starting from the coarsest and ending in the finest scale, the next 8 tensors encode the $v_x$ dimension, starting from the finest and ending in the coarsest scale, and the final $9$ tensors encodes the $x$ dimension, starting from the coarsest and ending in the finest scale. This encoding can be seen pictorially in Fig.~\ref{fig:Beat_encodings} $a$. 

Our comb geometry consists of four branches numbered 1-4, each encoding a separate dimension, namely $y,v_y,x$ and $v_x$ respectively. Each branch on its own is simply an MPS, however we connect these branches together along the first tensors of each branch, as illustrated in Fig.~\ref{fig:Beat_encodings} $b$. These bonds which connect between the different dimensions are collectively known as the spine of the comb network. Different encodings and number of branches were tested but we found this as the choice which required the lowest numerical resources for the beat wave. We consider a case where each tensor carries a physical index, giving a comb geometry with 32 tensors for our simulations. The process of generating a comb tensor network is well understood where once can convert an MPS into a comb via a process of SVDs, the details of which are outlined in Appendix \ref{App:comb}. 

We use standard RK4 to propagate the distribution functions and EM fields in time, using a fixed time-step of $\Delta t= 0.002$, where all spatial derivates are evaluated using central finite-difference schemes. In order to simulate the electric fields we split apart the fields from the lasers $\mathbf{E_L}$ and those generated from the plasma $\mathbf{E_p}$. These separate fields are then propagated in time, according to,
\begin{eqnarray}
    \frac{\partial \mathbf{E_p}}{\partial t} = \nabla \times \mathbf{B_p} - \mathbf{J} \ \ , \frac{\partial \mathbf{E_L}}{\partial t} = \nabla \times \mathbf{B_L} \ , \label{eq:EL_dt}\\ 
     \frac{\partial \mathbf{B_p}}{\partial t} =- \nabla \times \mathbf{E_p} \ \ , \frac{\partial \mathbf{B_L}}{\partial t} =- \nabla \times \mathbf{E_L} \ . \label{eq:Elp_dt} 
\end{eqnarray}
The total Electric field $\mathbf{E} = \mathbf{E_L}+\mathbf{E_p}$ is then computed, which is used to perform the temporal evolution of the plasma according to Eq.~\eqref{eq:fluc}. The incoming lasers are injected into either end of the plasma, and freely outflow after reaching the opposite end, i.e. we consider continuous boundary conditions at the opposite end from injection. We outline how one can inject waves and allow for outflow using finite differences and tensor networks in Appendix \ref{App:inject}.

\subsection{Comparisons between tensor network geometries}
\begin{figure}[htb!]
    \centering
    \includegraphics[width=\linewidth]{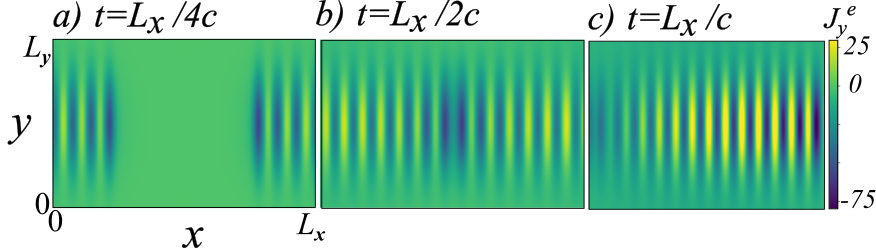}
    \caption{Snapshots of current density from the electron distributions in the plasma at various times. $a)$ Current density at a time before the lasers intersect, $b)$ Current density as the lasers meet at the centre and $c)$ Current density after beat wave has developed. }
    \label{fig:Jy_snaps}
\end{figure}

\begin{figure}[thb!]
    \centering
    \includegraphics[width=\linewidth]{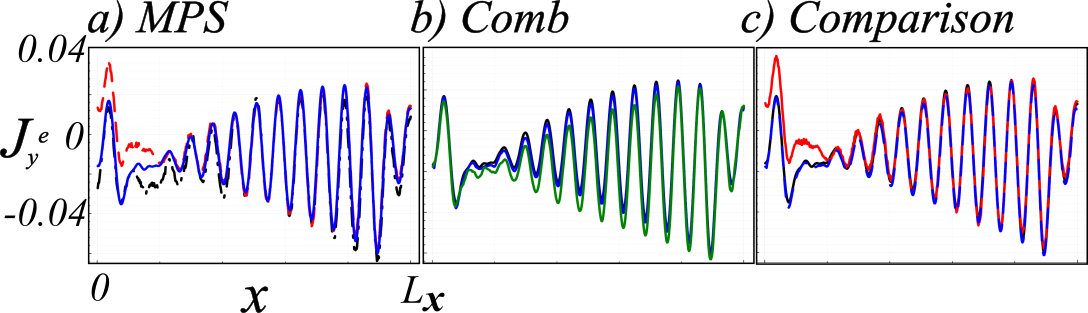}
    \caption{Demonstration of convergence of results for MPS and comb simulations. We plot a cut along the x-y plane at $y=L_x/2$ of the current density from the electrons, $J_y^e(x,y)$ for the different tensor network geometries at a fixed simulation time of $t=0.14$. We demonstrate how well the dynamics converge as one increases the bond dimension of the distribution function $\chi_f$ and the intermediate derivative $\chi_K$. $a)$ Convergence of the MPS simulations for bond dimensions of $(\chi_f,\chi_K)=(150,250)$ (dash doted black), $(\chi_f,\chi_K)=(200,250)$ (dashed red) and $(\chi_f,\chi_K)=(512,250)$ (solid blue). $b)$ Convergence of the comb simulations for bond dimensions of $(\chi_f,\chi_K,\chi_S)=(128,128,32)$ (green), $(\chi_f,\chi_K,\chi_S)=(128,128,64)$  (blue) and $(\chi_f,\chi_K,\chi_S)=(128,128,128)$ (black). In $c$) we plot the output from the MPS simulations of $\chi_f=200$ (solid red) and $\chi_f=512$ (solid black ) along with the comb simulation with $(\chi_f,\chi_K,\chi_s)=(128,128,64)$ (dashed blue) as a comparison.  }
    \label{fig:Jy_conv}
\end{figure}

The plasma is initialised as outlined in the previous subsection from which we simulate the full propagation of the incoming laser beams from the boundaries at $t=0$. The laser electric fields are polarised along the $y$ axis leading to a propagation horizontally across the $x$ axis.  The electric field from each laser causes an acceleration of the charges particles in the plasma, leading to a current to develop as the different ionic species respond to the incoming oscillating electric fields. We demonstrate snapshots of the current density along the $y$ axis caused by the electrons, defined as
\begin{equation}
    J_y^e(x,y,t)= \iint v_y f_e(x,y,v_x,v_y) dv_{x} dv_{y} ,
\end{equation}
in Fig.~\ref{fig:Jy_snaps} at various times in the simulation. One observers clear oscillatory dynamics in the current density in line with the incoming lasers. After the lasers overlap at a time $t=L_x/2c$ the beat wave begins to develop, which can be observed in the current density of the electrons in Fig.~\ref{fig:Jy_snaps}$c$.
We perform identical simulations with matching parameters using both the MPS and comb tensor network geometries across various bond dimensions and compare the resultant dynamics and representibility of the distribution function. As we are utilising an RK4 evolution scheme at each time-step we have $f_s(x,v,t)$ encoded as an MPS/comb and also the intermediate derivatives $K=\frac{\partial f}{\partial t}$. We truncate the tensor networks representing the $f$ states to a maximal bond dimension of $\chi_{f}$, and truncate the intermediate derivative states $K$ to a bond dimension of $\chi_K$. We first compare the resultant current densities $(J_y^e)$ produced from each geometry. In Fig.~\ref{fig:Jy_conv} we plot a slice of the current density $J^e_y(x,y)$ at a fixed simulation time of $t=0.14$ along the midpoint of the y axis, $y=L_y/2$ for simulations with different bond dimensions. All simulations are initialised with the same conditions and thus the discrepancies in Fig.~\ref{fig:Jy_conv} arise from making use of varying bond dimensions. Focusing first on the MPS simulations presented in Fig.~\ref{fig:Jy_conv}$a$, we truncate our intermediate derivative states to a bond dimension of $\chi_K=250$ and vary the bond dimension of the distribution function $f$ $\chi_f$. We observe that all bond dimensions used are able to capture the correct period of oscillations in the beat wave, with agreement between the placement of the peaks and troughs of the wave. However there is still substantial deviations between the lower bond dimensions and the $\chi=512$ run, particularly where the beat wave flattens out in $x \in [\frac{L_x}{8},\frac{L_x}{4}]$. In this region we can clearly observe the presence of noise in the lower bond dimension runs, and noticeable discrepancies between all the runs which indicates that the simulations of $\chi=150$ and $\chi=200$ have not converged yet, even at these large bond dimensions, demonstrating the slow convergence of the MPS results.

Considering next the comb geometry we truncate both $f$ and $K$ states to different maximal bond dimensions, however we choose also to truncate the spine of the comb further to a bond dimension of $\chi_S$. We see remarkable agreements between all bond dimensions used in the simulations from the largest run of $(\chi_f,\chi_K,\chi_S)=(128,128,128)$ down to the more truncated runs of $(128,128,64)$, $(64,128,64)$ and $(64,64,64)$. Across all runs, the period of the beat wave remains consistent, as with the MPS case, however there is very little discrepancy even at the troublesome region for MPS of $x \in [\frac{L_x}{8},\frac{L_x}{4}]$. Even the most heavily truncated simulation of $(\chi_f,\chi_K,\chi_S)=(128,128,32)$ agrees with the other simulations and actually looks to out-perform the MPS run of $(\chi_f,\chi_K)=(200,250)$. The only signs of truncation are slightly lower amplitudes for the peaks of the beat oscillation for the more heavily truncated runs. Contrasting these results against the MPS case demonstrates that by using the comb geometry one can converge the resultant dynamics with fewer numerical resources.

\begin{figure}[thb!]
    \centering
    \includegraphics[width=.9\linewidth]{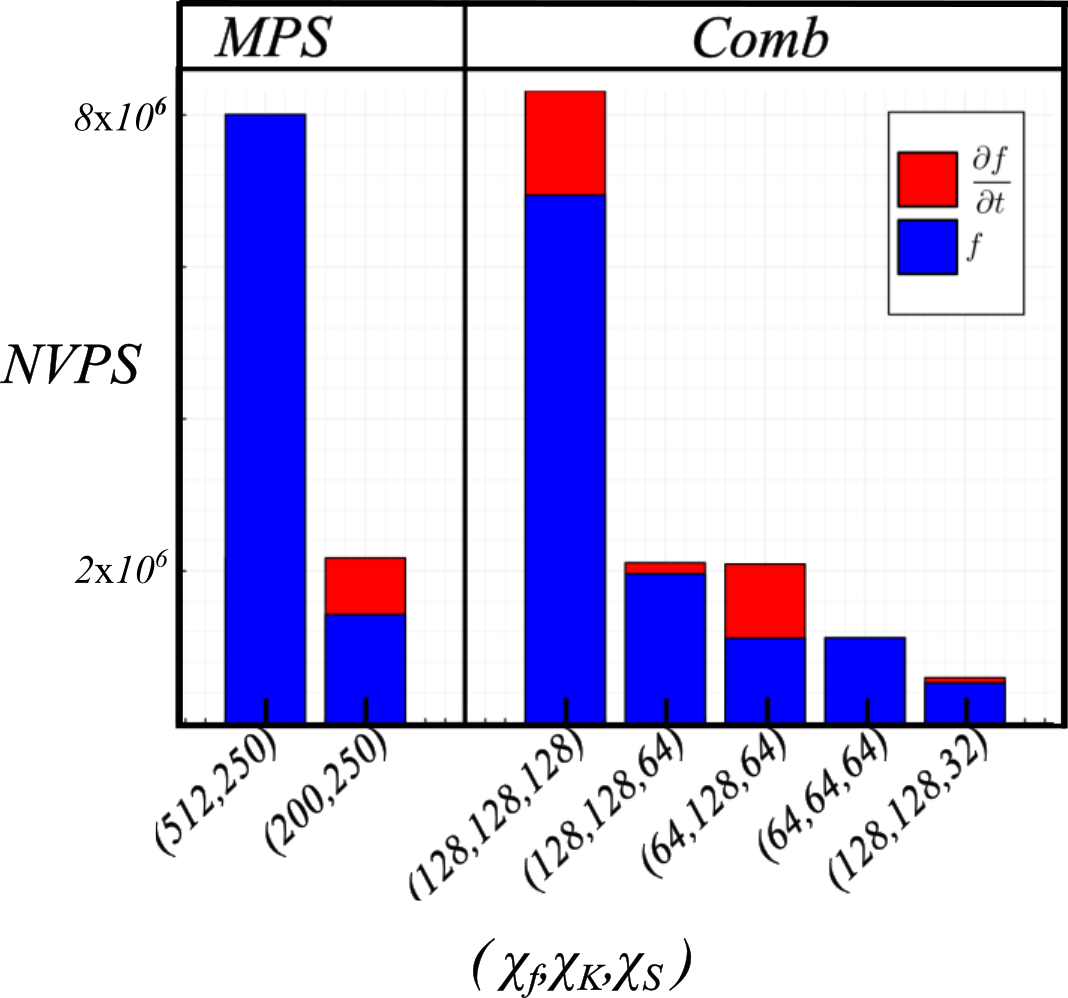}
    \caption{Comparison of the total number of variational parameters (NVPS) inside each tensor network considered. We plot the NVPS required to represent the plasma distribution $f$ (blue) and the intermediate derivatives $\frac{\partial f}{\partial t}=K$ (red) for the MPS (first two bars) and the comb simulations (final five bars). We show NVPS across different maximal bond dimensions where $\chi_f$ is bond dimension to represent $f$, $\chi_K$ to represent $K$ and $\chi_S$ is the spine bond dimension for the comb geometry (which does not exist for the MPS simulations). The convergence of the dynamics between the comb geometry and the $\chi_f=512$ MPS simulation suggests that one can converge the dynamics with fewer numerical resources using the comb geometry. }
    \label{fig:Beat_NVPS}
\end{figure}

In Fig.~\ref{fig:Jy_conv}$c$ we overlay the largest $\chi$ simulations of the MPS with the truncate comb run of $(\chi_f,\chi_K,\chi_S)=(128,128,64)$, and observe near perfect agreement between these runs, indicating a clear convergence between the large MPS simulations and the comb geometry.  However overlaying the $\chi_f=200$ MPS result shows substantial deviation even to the truncated comb runs. This indicates that the results we obtain from a heavily truncated comb tensor network simulation are comparable to a much larger bond dimension MPS simulations. One should point out that as these are two different geometries, the resultant errors from truncation may have different effects on the resultant simulations. It is also important to compare the total number of parameters (NVPS) needed to describe the two tensor networks, with a lower NVPS representing a higher degree of data compression. We plot the NVPS for different truncation levels of the MPS and comb networks in Fig.~\ref{fig:Beat_NVPS}, and observe that the $(\chi_f,\chi_K,\chi_S)=(128,128,128)$ comb simulation requires on the order of $8\times10^6$ numerical parameters which is just as many parameters as the $\chi_f=512$ MPS simulation. Truncating the MPS to $\chi_f=200$ reduces this to $2\times10^6$ parameters, which is equal to or larger than the number of parameters required for the rest of the considered comb simulations, all of which produced comparable results to the $\chi_f=512$ MPS run. As such one can conclude that the comb geometry, for this particular choice of beat wave parameters, requires far fewer numerical resources to converge as compared to the MPS case.
Furthermore, these results reflect that the improved accuracy of comb results over MPS is arising from the increased connectivity of the comb geometry and not simply due to the increased number of parameters in the comb description compared to an equivalent bond dimension MPS.

\begin{figure}[htb!]
    \centering
    \includegraphics[width=\linewidth]{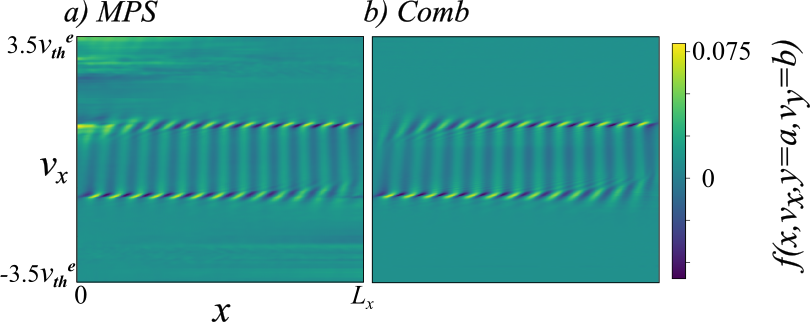}
    \caption{Snapshot of the dynamics of $f_e$ in the $x-v_x$ plane for a fixed $y=a$ and $v_y=b$. We show the resultant 2D distribution function one obtains at a simulation time of $t=L_x/c$ when one uses $a)$ an MPS of bond dimension $\chi=200$ and $b)$ a comb with a bond dimension $\chi=64$. We chose to take a slice in phase space along the midpoint of the $y$ dimension ($a=L_y/2$) and a small positive $v_y$ ($b=0.125 v_{max}$). One can observe apparent noise in the MPS simulation relative to the comb via comparing the top left and bottom right corners of $a)$ and $b).$}
    \label{fig:xvx}
\end{figure}

We additionally plot a 2D slice through phase space in Fig.~\ref{fig:xvx} from the $\chi_f=200$ MPS simulation and the $(\chi_f,\chi_K,\chi_S)=(128,128,64)$ comb simulation. We find a detailed structure builds in phase space, with a complicated oscillation developing in $v_x$ as one varies $x$. Although both these MPS and comb simulations agree on the general features in phase space, we can more clearly observe the noise in the MPS simulations towards the edges of simulations, which is not present in the comb simulations, or at least not present to the same level.

\begin{figure}[htb!]
    \centering
    \includegraphics[width=.75\linewidth]{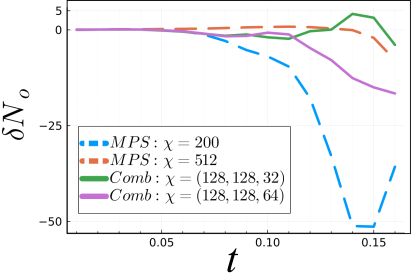}
    \caption{Demonstration of convergence of results for MPS and comb simulations. We plot the total number fluctuation of the number of plasma particles in the plasma. Simulations performed with MPS are illustrated with dashed lines, whilst those form the comb geometry are represented by solid lines. For both geometries, we plot the behaviour of the number fluctuations as a function of time and for different bond dimensions $\chi$. The total number of particles is a conserved quantity for Vlasov evolution and thus a perfect simulation will produce $\delta N_o=0$ at all times.}
    \label{fig:No_conv}
\end{figure}

As a further metric of success for each of the tensor network simulations one can calculate additional observables of the plasma dynamics such as the total number of particles. By plotting such observables as a function of time for various bond dimensions, one can identify how well the simulations have converged. We plot the fluctuations of the number of electrons obtained from,
\begin{equation}
    \delta N_o = \iiiint \tilde{f}_e(x,y,v_x,v_y) dx dy dv_{x} dv_{y}  \ . 
\end{equation}
In the Vlasov system the total particle number is a conserved quantity and thus $\delta N_o=0$ for all times in a perfect simulation. Looking at the number fluctuations in Fig.~\ref{fig:No_conv} we see that the MPS runs with $\chi_f=200$  produce fluctuations on the order of $|\delta N_o|\sim50$ at a simulation time of $t=0.15$. In contrast the most heavily truncated comb geometry yields $|\delta N_o|\sim10$ at the same simulation time. Increasing the MPS to the more computationally demanding $\chi_f=512$ does reduce this fluctuation to $|\delta N_o| \sim 6$, however the comb run of $(\chi_f,\chi_K,\chi_s)=(128,128,32)$ and $(128,128,64)$ are able to roughly match this performance.
We fully expect that for both the MPS and comb geometries that increasing the bond dimensions further, in addition to decreasing the time and spatial discretisation errors inherent in a finite difference RK4 time evolution scheme, would be able to further reduce $\delta N_o$ across all times.

Across all the results for the beat wave problem presented here, we can therefore confirm that the comb geometry can indeed aid in tensor network simulations compared to the simpler MPS geometry. Although both tensor networks can produce successful simulations of plasma dynamics for sufficient bond dimensions, the comb geometry lends itself to requiring a noticeably smaller bond dimension compared to the MPS to produce accurate simulations, reducing the computational resources required.

\subsection{Validation against PIC code}
\begin{figure}[t]
    \centering
\includegraphics[width=\linewidth]{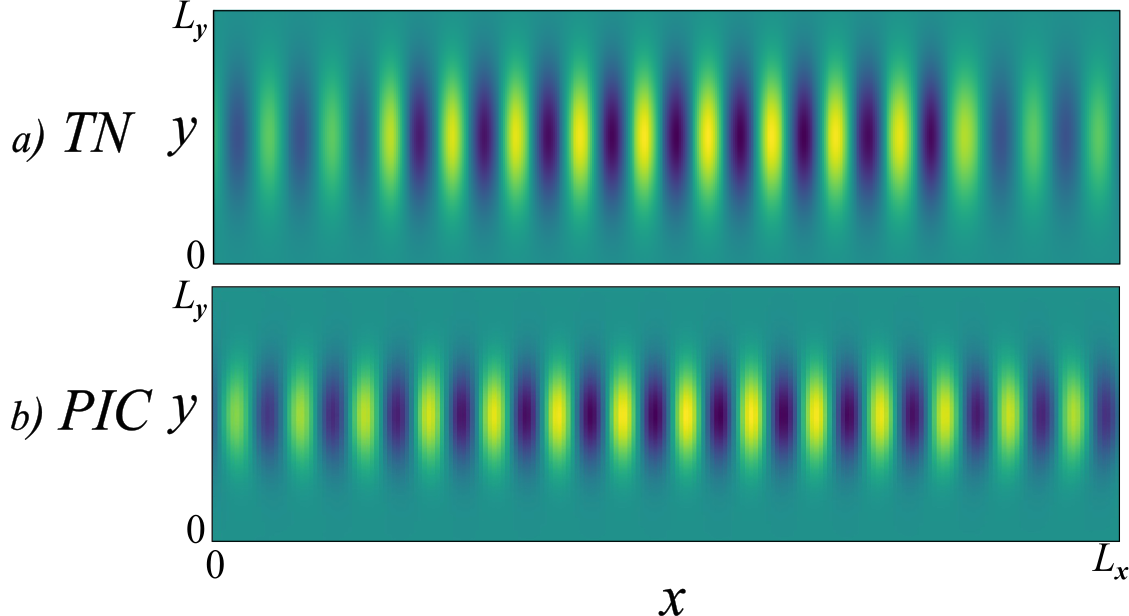}
    \caption{Comparison between dynamics of the $E_y$ field from $a)$ the tensor network (TN) and $b)$ the Particle In Cell (PIC) implementation of the beat wave problem at a simulation time of $t=0.125$. For our tensor network results, we are using a comb geometry with bond dimensions of $(\chi_f,\chi_K,\chi_S)=(128,128,64)$.
    }
    \label{fig:PIC}
\end{figure}

As a sanity check the exact beat wave setup was also constructed and simulated using a PIC code implemented via the Chicago software \cite{PIC_beat,PIC_gas,Beat_paper}. We show snapshots of the dynamics of the Electric fields in Fig.~\ref{fig:PIC} from both the PIC code and our tensor network approaches. We observe close qualitative agreement between the two approaches, with the Electric field snapshot showing a wave profile with matching periodicity and overall beat envelope. There are some noticeable differences between the two approaches however. Most notably in the PIC simulations the Electric field phase fronts show a small degree of bending towards the $x$ boundaries which we do not capture with our tensor network implementation. This is due to the PIC codes implementing a full Gaussian beam propagation for the injected lasers, however in our tensor network implementation instead of a full Gaussian beam simulation, we inject plane waves at either boundary and multiply by a Gaussian envelope afterwards, as outlined in Appendix \ref{App:inject}. This difference in the implementation of the injected lasers may account for the differences observed, but overall we still observe qualitative agreement between our tensor network simulations and those from the PIC simulations.

\section{Magnetohydrodynamics} \label{sec:MHD}

Turning attention away from the Vlasov-Maxwell system of equations, we now consider how tensor networks could be applied to the alternative treatment of plasma dynamics in the form of Magnetohydrodynamics (MHD). As opposed to the fully kinetic phase space approach of the Vlasov equation, the MHD formalism treats the plasma as a compressible fluid in real space ($\mathbf{x}$) with a plasma velocity $\mathbf{u}(\mathbf{x})$ and corresponding density $\rho(\mathbf{x})$, pressure $p(\mathbf{x})$ and internal energy $E(\mathbf{x})$, along with the self generated magnetic field $\mathbf{B}(\mathbf{x})$. The simplest form of MHD is known as ideal MHD and the governing equations are as follows,
\begin{eqnarray}
    &\frac{\partial \rho}{\partial t} = -\nabla \cdot (\rho \mathbf{u})  \ , &\\
    &\frac{\partial \rho \mathbf{u}}{\partial t} = -\nabla \left[ \rho \mathbf{u}\mathbf{u}^T + (p+ \frac{1}{2}B^2 ) I_{3\times3} -\mathbf{B}\mathbf{B}^T \right]  \ , &\\ 
  &  \frac{\partial E}{\partial t} = -\nabla \cdot \left[  (\frac{\gamma}{\gamma-1} p +\frac{1}{2}\rho u^2) \mathbf{u} - (\mathbf{u}\times\mathbf{B})\times \mathbf{B} \right] \ , & \\
  &\frac{\partial \mathbf{B}}{\partial t} = \nabla \times (\mathbf{u}\times \mathbf{B})  \ . &
\end{eqnarray}
In the above expression $I_{3\times3}$ corresponds to a $3\times3$ Identity matrix, and $\gamma$ is a constant defined as the ratio of the specific heats $\gamma=\frac{C_p}{C_v}$, with $C_p$ and $C_v$ the specific heat capacity at constant pressure and volume respectively. One additionally requires an equation of state to relate the pressure to the internal energy which is taken as,
\begin{equation}
    p=(\gamma -1) (E-\frac{1}{2}\rho \mathbf{u}^2 -\frac{1}{2}\mathbf{B}^2) \ .
\end{equation}

A major benefit of dealing with the MHD formalism is that one works entirely in position space, as opposed to the Vlasov description where the velocity is considered as a separate dimension which must be encoded into the tensor network. The trade-off is that for MHD one must simultaneously solve more coupled equations. Encoding MHD into tensor networks follows a similar procedure to that outlined previously. Each of $\rho$, $E$, $p$, and each component of $\mathbf{u}$ and $\mathbf{B}$ is encoded as a separate MPS, which can then be evolved via standard time marching schemes. 

\subsection{Orszag-Tang Vortex}
As a first demonstration of using tensor networks for MHD we consider the standard test case of the Orszag-Tang vortex system \cite{Orszag_Tang_1979,DAI1998331}. The Orszag-Tang vortex can equally be treated via the kinetic Vlasov system, and previous work has successfully explored using tensor networks in the form of a comb geometry to model this problem \cite{Plasma_comb}. Here we demonstrate that by using the MHD formalism we can equally well simulate this test case and that we can do so whilst still exploiting the simplest tensor network in the form of MPS. The Orszag-Tang vortex is a 2D problem which models the dynamics of a plasma subject to the following initial conditions,
\begin{eqnarray}
   & \rho =\gamma ^2 \ \ , \ \ u_x = -\sin(y) \ \ , \ \ u_y =\sin(x) \ ,& \\ 
  &  p =\gamma  \ \ , \ \ B_x = -\sin(y) \ \ , \ \ B_y =\sin(2x) \ .&
\end{eqnarray}
We take $\gamma=\frac{5}{3}$ throughout, with periodic boundaries in all directions. As this is a 2D problem, the full set of equations one must solve can be expressed as,
\begin{widetext}
\begin{equation}
    \frac{\partial}{\partial t} \begin{pmatrix}
        \rho \\ \rho u_x \\ \rho u_y \\ B_x \\ B_y \\ E 
    \end{pmatrix} =- \frac{\partial}{\partial x} \begin{pmatrix}
        \rho u_x \\  \rho u_x^2 +p^*-B_x^2 \\ \rho u_x u_y -B_x B_y \\ 0 \\ B_y u_x -B_x u_y \\ (E+p^*)u_x - B_x (u_x B_x +u_y B_y)
    \end{pmatrix} - \frac{\partial}{\partial y} \begin{pmatrix}
         \rho u_y \\ \rho u_x u_y -B_x B_y\\ \rho u_y^2 +p^*-B_y^2 \\ B_x u_y -B_y u_x \\ 0 \\ (E+p^*)u_y - B_y (u_x B_x +u_y B_y)
    \end{pmatrix}  \ ,
\end{equation}
\end{widetext}
where $p^*$ is known as the total pressure and is given by $p^*=p+\frac{1}{2}\mathbf{B}^2$. These initial conditions evolved under the ideal MHD equations generate a central vortex, with outward propagating shock waves which compress the plasma. When solving the MHD equations, we evolve the momentum $\rho u_x$ and $\rho u_y$ as separate MPS, however we need explicitly the velocities $u_x$ and $u_y$ to calculate all of the derivative terms present in the ideal MHD formalism. To perform this effective division of two tensor networks we exploit a variational scheme to obtain an MPS representation for the inverse density $\frac{1}{\rho}$, where one performs an optimisation over an initial guess MPS at every time step. We highlight the details of this method in Appendix \ref{app:inverse_den}. We choose to perform the Orszag-Tang simulation using the $4th$ order Runge-Kutta scheme, where after each time-step we apply a Shuman filter \cite{Shapiro} in order to suppress the smallest scale numerical oscillations and increase the stability of the code. One can adaptively control the level of filtering via a parameter $\beta$, where the smaller the $\beta$, the larger the filtering effect and thus the more smoothing of the function is implemented. Further details of the filtering process are laid out in Appendix \ref{App:Shuman}, and unless otherwise stated we use a filtering parameter of $\beta=5$.  We perform our simulations on a $2^9\times 2^9$ spatial grid resulting in an 18 site MPS to encode each of the dynamical variables. During these simulations, each of the $6$ MPS representing the dynamical variables are propagated in time, consisting of a series of MPO-MPS contractions and additions of MPS. These operations are performed using a relative truncation cutoff of $10^{-15}$, whilst also imposing a maximal possible bond dimension $\chi$ for all MPS in the simulation. This allows the bond dimensions of each MPS to individually grow dynamically during the simulations up to the specified maximal bond dimension.

\begin{figure*}[t]
    \centering
     \includegraphics[width=\linewidth]{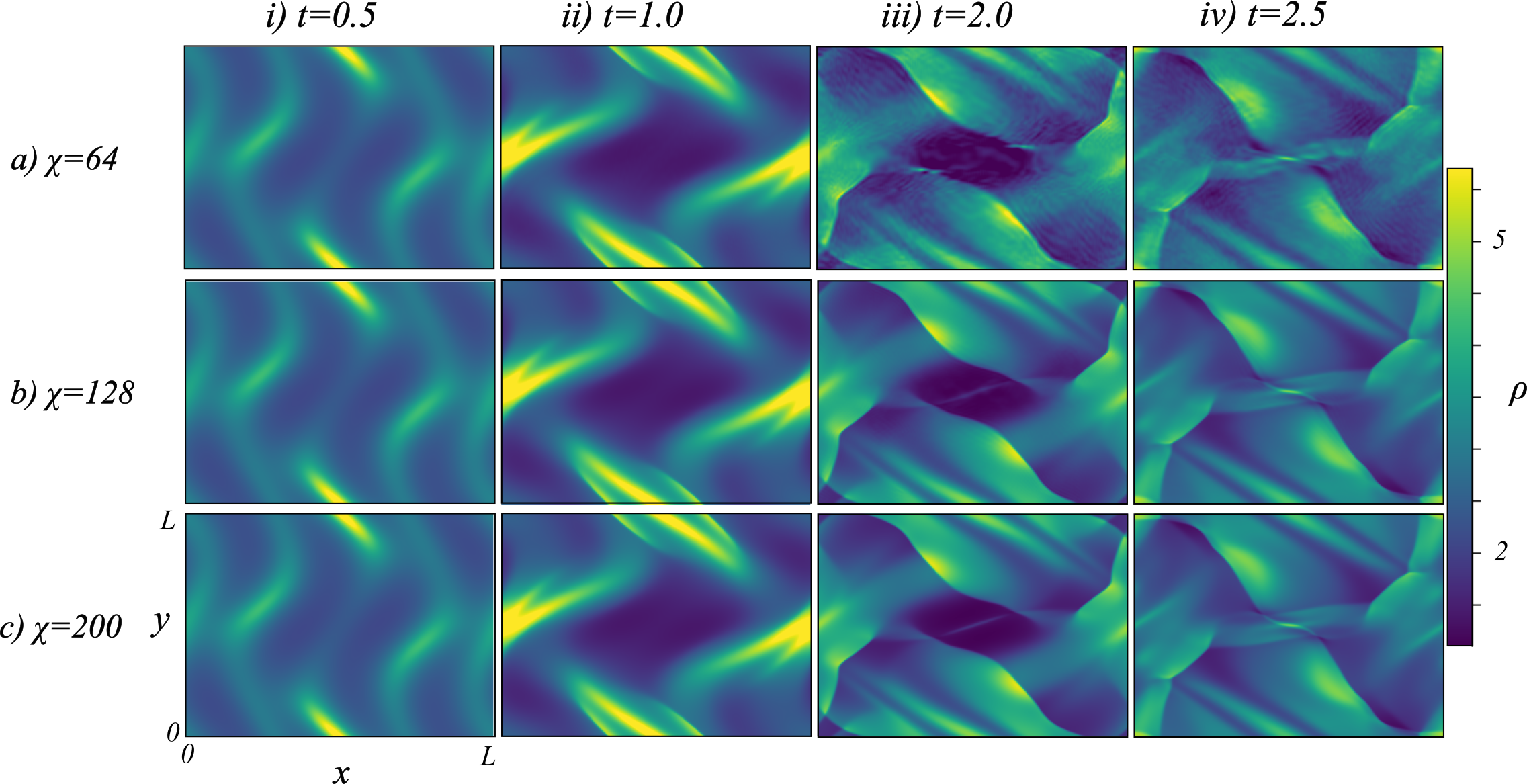}
    \caption{Time evolution of the density $\rho$ in the Orszag-Tang vortex on a $M_x \times M_y =2^9 \times 2^9 $ spatial grid, where $x,y\in[0,L]$ with the length of the box taken as $L=2\pi$. We perform the time evolution using a fourth order Runge-kutta scheme with a spatial filter, with a fixed time-step of $dt=dx/5$, whilst using maximal bond dimensions of $a)\chi=64$, $b) \chi=128$, and $c)$ $\chi=200$, and a truncation cutoff of $10^{-15}$ throughout. }
    \label{fig:MHD_snapshots}
\end{figure*}

\begin{figure*}[t]
    \centering
     \includegraphics[width=\linewidth]{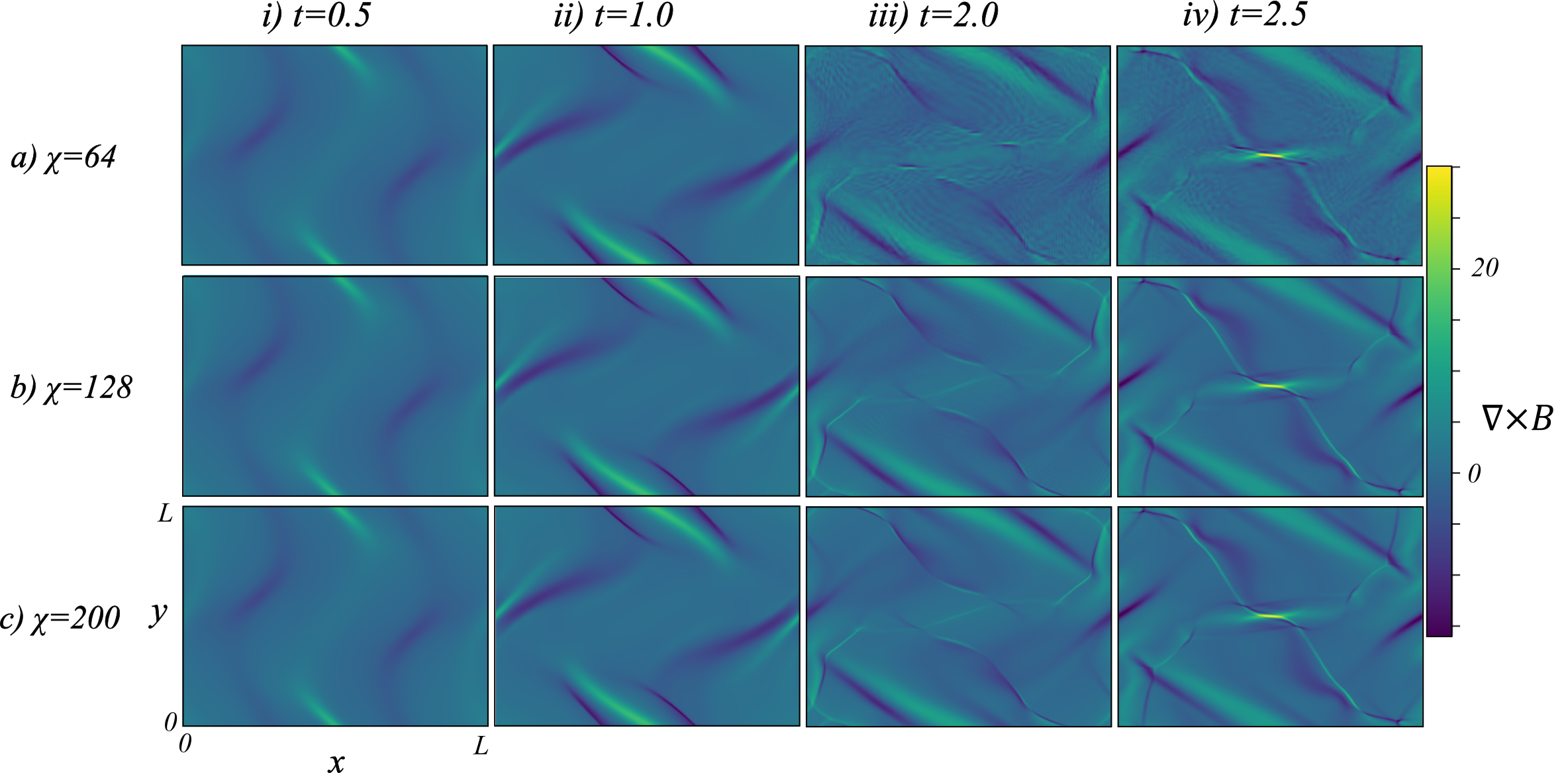}
    \caption{Time evolution of the curl of the magnetic field $\nabla \times \mathbf{B}$ in the Orszag-Tang vortex on a $M_x \times M_y =2^9 \times 2^9 $ spatial grid, where $x,y\in[0,L]$ with the length of the box taken as $L=2\pi$. We perform the time evolution using a fourth order Runge-kutta scheme with a spatial filter, with a fixed time-step of $dt=dx/5$, whilst using maximal bond dimensions of $a)\chi=64$, $b) \chi=128$, and $\chi=200$, and a truncation cutoff of $10^{-15}$ throughout. }
    \label{fig:MHD_Bz_snapshots}
\end{figure*}

Snapshots of the resultant dynamics are shown in Fig.~\ref{fig:MHD_snapshots} for the density of the plasma $\rho$, and we display the curl of the magnetic field $\nabla \times \mathbf{B}$ in Fig.~\ref{fig:MHD_Bz_snapshots} for various maximal bond dimensions $\chi$. We observe clear qualitative agreement between the dynamics at bond dimensions of $\chi=64, 128$, and $200$, with each simulation showing the formation of the central vortex and emergence of shock waves, which are identified as jumps in the density distribution. Looking first at the density, no difference can be found in the simulation between a bond dimension of $\chi=128$ and $\chi=200$, indicating clear convergence.  In contrast, although the $\chi=64$ simulation does produce the qualitatively correct dynamics, there is a substantial amount of noise present, particularly at later simulation times of $t\ge2$. By heavily truncating the MHD system at each time-step, truncation errors are introduced into the simulation which can build over time leading to a substantial amount of noise at later times. Additionally as one evolves the Orszag-Tang test case we expect to develop more and more correlations within the MPS which requires larger bond dimensions to properly represent. This is also reflected in the snapshots of the curl of the magnetic field in Fig.~\ref{fig:MHD_Bz_snapshots}, where again the $\chi=64$ simulation shows an abundance of noise, which is not present in the $\chi=128$ and $\chi=200$ simulations.  

\begin{figure}[t]
    \centering
     \includegraphics[width=1.0\linewidth]{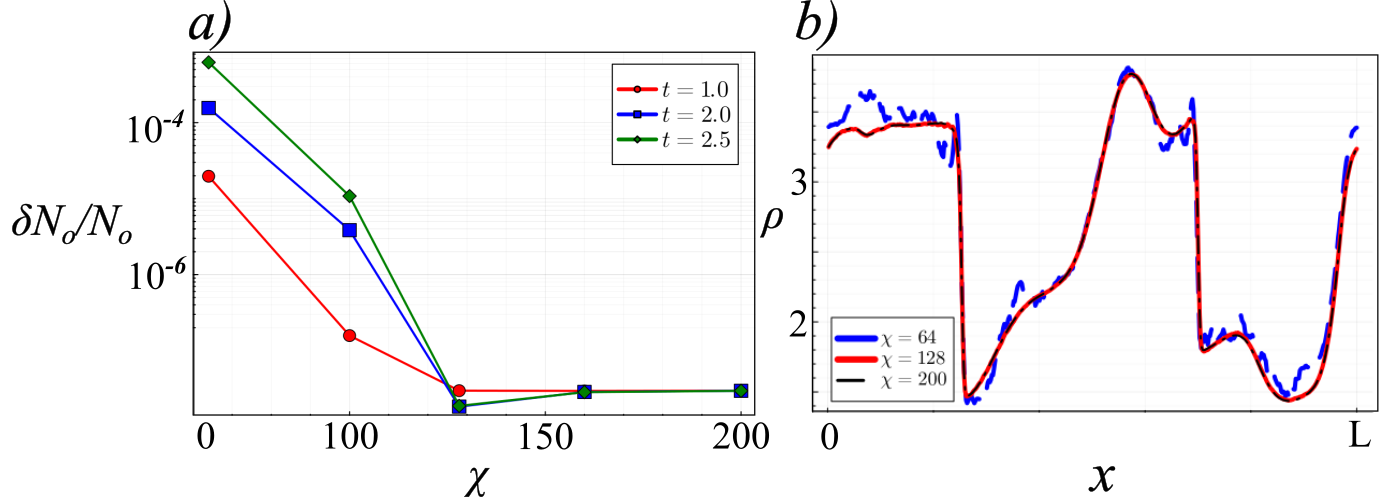}
    \caption{$\mathrm{a})$ Plot of the change in total number of plasma ions in the test case as a function of bond dimension $\chi$, obtained via integrating the density over all space and subtracting from the original number of plasma ions at $t=0$. As this is a conserved quantity, for a perfect simulation $\delta N_o=0$. We compare this quantity for simulation times of $t=1.0$ (red dots) , $t=2.0$ (blue squares), and $t=2.5$ (green diamonds). $\mathrm{b})$ Cut across the $y=1.95$ plane at a simulation time of $t=3.0$ for simulation bond dimensions of $\chi=200$ (dash dot black), $\chi=120$ (solid red) and $\chi=64$ (dashed blue). Note that the $\chi=200$ and $\chi=128$ lines appear to lay directly over one another.  }
    \label{fig:MHD_No_cons}
\end{figure}

As an additional benchmark, we compute the integral over the density $\rho$ as a function of maximal bond dimension $\chi$ for various simulation times in Fig.~\ref{fig:MHD_No_cons}, which provides a measure for the total number of ions in the plasma, $N=\iint\rho(x,y)dxdx$. This is a conserved quantity, and thus by measuring the fluctuation of the number of ions as a function of time, $\delta N_o= \iint\rho(x,y,t)-\rho(x,y,t=0) dxdy$, one can assess how well a simulation respects this condition, with the expectation that severely truncating an MPS will have a negative impact on the conservation. It should be pointed out however that truncation of the MPS is not the only cause of a non-zero $\delta N_o$, the spatial and temporal discretisation also play a contributory role. Across all simulation times one can reduce $\delta N_o$ simply by increasing our maximal bond dimension up to $\chi =128$. Increasing $\chi$ beyond this does not produce any improvement upon the fluctuation of particle number. This indicates that beyond $\chi=128$ the dominant source of error for the total particle number does not lay within truncation of our MPS, but instead in the spatial or temporal discretisation, or in the application of the filter. In Fig.~\ref{fig:MHD_No_cons}$b)$ we also display the density at a cut through a shock wave on the $y=1.95$ plane. We observe that at all maximal bond dimensions we can resolve the shock wave formation, whilst over truncating the bond dimension still allows one to identify the rough features of the shock wave, i.e. the presence and location of the shock, however with noticeable noise present in the form of small scale oscillations in the density close to the shock fronts.

Based on these results we see that MPS in this regime is able to capture the overall Orszag-Tang dynamics, resolving shock formation with agreement against the expected results in literature, and demonstrating convergence with increasing bond dimensions, whilst providing data compression. Using $\chi=64$ requires only storage of $23\%$ of numerical data compared to that of a direct numerical simulation on a likewise fixed grid which rises to $66\%$ for $\chi=128$. Although the dynamics can be captured using MPS without relying on more complex geometries, one does need to evolve many more coupled equations in the MHD approach as compared to the Vlasov approach \cite{Plasma_comb}. The MHD model can be applied beyond this Orszag-Tang test case and is well suited for the modelling of plasmas where the dynamical time-scales are long relative to the microscopic dynamics of the ions, and that the spatial scales are large relative to the Debye length, defined as $\lambda_D^2=\frac{\epsilon_0 k_B T}{n q^2}$ with $T,n,q,\epsilon_0$ and $k_b$ the temperature, density, unit charge, electric permittivity of free space and Boltzmann constant respectively.




\section{Conclusion}

We have demonstrated the ability for tensor networks to perform a variety of plasma simulations and shown the application to industrially relevant problems. Although MPS simulations of the Vlasov equations can perform well for certain examples, we have identified that as one introduces further complexity in form of strong magnetic fields or complicated laser interactions, one may need to consider alternative approaches. MPS are the simplest form of tensor networks and thus extending to more complex geometries which can better handle the growth of correlations within the plasma is a natural path. We have confirmed that the comb tensor network allows one to perform simulations of a laser plasma interaction with bond dimensions smaller than those required from MPS for the same accuracy level. Using a more complex tensor network geometry can lead to a more numerically intensive calculation however, as in the comb geometry one must operate on tensors with three bond indices instead of just two for MPS, leading to an extra factor of $\chi$ in the computational cost scaling for the comb geometry \cite{Plasma_comb}. 

In some cases one can consider plasmas via the MHD formalism which is widely used for the simulation of stellar formation. In the regimes considered here it appears that an MPS encoding is sufficient to produce reliable dynamics, capturing the relevant physics of shock formation whilst still providing data compression, without the need to consider a more complex tensor network. Simulations of the MHD equations does require one to simultaneously evolve many more coupled PDEs as opposed to the single Vlasov equation, and further requires consideration of how to best perform the division of two tensor networks.

Future work may explore if alternative tensor network geometries such as a tree tensor network \cite{TTN_miles_func,TTN_gen_model,TTN_qcircs}, or variations upon the comb geometry can produce an additional benefit for simulations of complex plasma dynamics and for other challenging PDEs. Additionally, thus far studies into applying tensor networks to the Vlasov-Maxwell system have focused on the collisionless regime \cite{Plasma_MPS,Plasma_comb}, where we neglect contributions due to the interactions between the plasma ions themselves. Future work may include additional interaction terms and analyse how well tensor networks are able to capture the resultant dynamics in regimes where collisions play an integral role in the dynamics \cite{doi:10.1098/rspa.2021.0167}. On the MHD formalism, we have thus far only considered ideal MHD \cite{MHD_notes}. Additional complexity can be added to the MHD equations to include more complex interactions such as the Hall effect \cite{hall_MHD} or a two-fluid description \cite{two_fluid_MHD}. 

\section*{acknowledgements}
We thank Bengt Eliasson, Archie Bott, Joe Gibbs, Neil Gaspar, and Dieter Jaksch for helpful discussions. This work was supported by the EPSRC through the Programme Grant QQQS (EP/Y01510X/1), and grant EP/Y005058/2. The authors would like to acknowledge AWE plc for financial support of the work.
UK Ministry of Defence © Crown owned copyright 2025/AWEs

\bibliographystyle{apsrev4-2}
\bibliography{main}


\appendix

\section{Generation of comb from MPS}\label{App:comb}
\begin{figure}[htb!]
    \centering
    \includegraphics[width=0.8\linewidth]{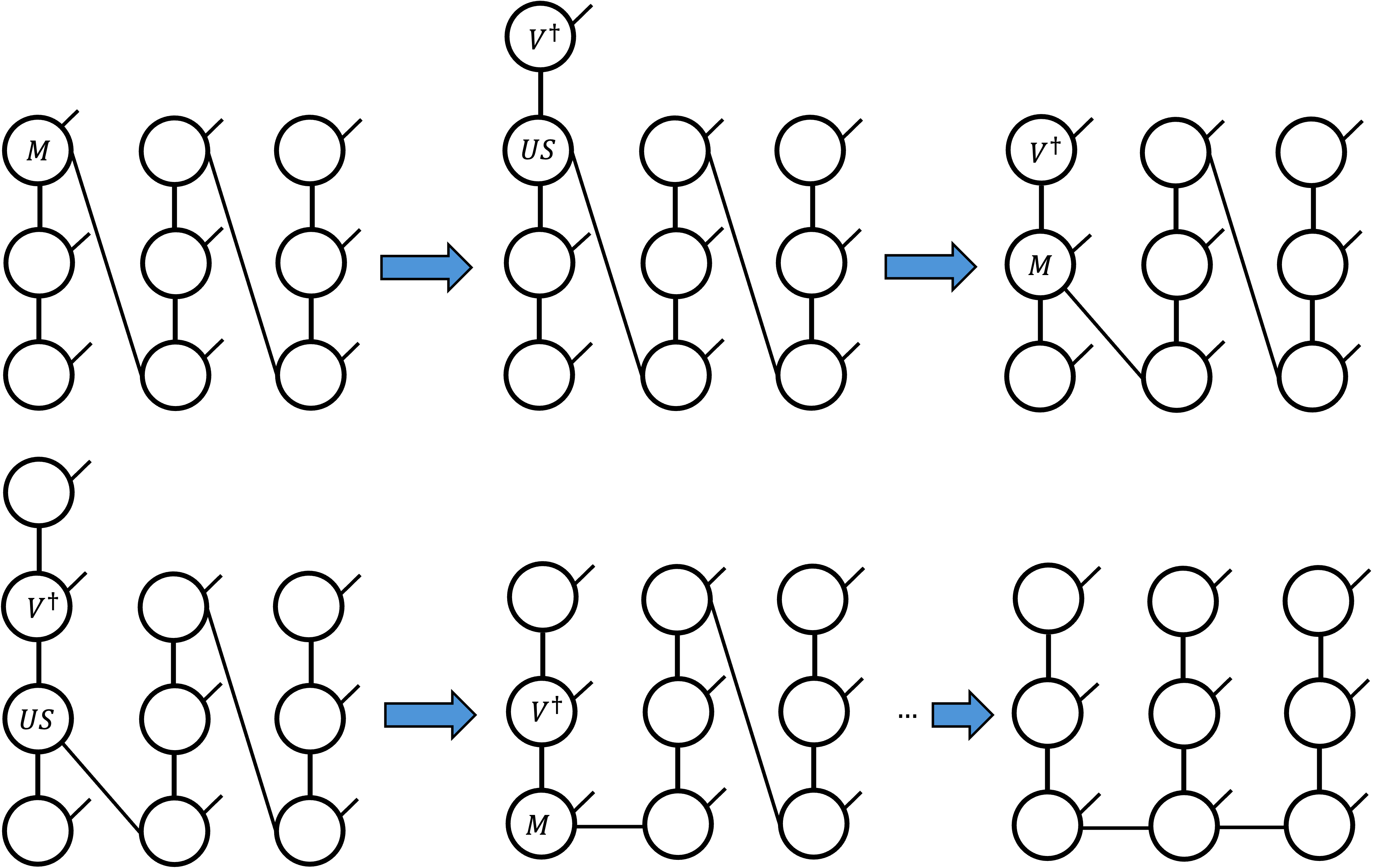}
    \caption{Process of converting an MPS into a comb tensor network. One begins with the tensor $M$ as shown in the top left of the diagram, and performs an SVD to push the linking bond between the first and second dimensions to the next tensor down the branch. One keeps performing SVDs to push this linking bond down to the first tensor in the first dimension. This is then repeated for the remaining dimensions until the comb geometry is formed. Note that although we have demonstrated this procedure using SVDs, one could equally perform the conversion via making use of QR decompositions instead }
    \label{fig:MPS_to_comb}
\end{figure}

Let us consider a three dimensional function $f(x,y,z)$ which is encoded into an MPS of three tensors per dimension,
\begin{widetext}
\begin{equation}
    f(x,y,z)= \sum_{\boldsymbol{\lambda}}A^{\sigma_1}_{1,\lambda_1}A^{\sigma_2}_{\lambda_1,\lambda_2}A^{\sigma_3}_{\lambda_2,\lambda_3}B^{\sigma_4}_{\lambda_3,\lambda_4}B^{\sigma_5}_{\lambda_4,\lambda_5}B^{\sigma_6}_{\lambda_5,\lambda_6}C^{\sigma_7}_{\lambda_6,\lambda_7}C^{\sigma_8}_{\lambda_7,\lambda_8}C^{\sigma_9}_{\lambda_8,1} \ ,
\end{equation}
\end{widetext}
where $\lambda_j$ denotes the bond indices and sites $1-3$ encode the $x$ dimension, $4-6$ encode $y$ and the final $7-9$ encode the $z$ dimension. To convert this MPS into a comb we wish to transfer the $\lambda_3$ bond from the $A^{\sigma_3}$ tensor to the $A^{\sigma_1}$ tensor which will become the spine of the comb. We begin by reshaping the $A^{\sigma_3}$ tensor and performing an SVD,  
\begin{widetext}
\begin{equation}
    f(x,y,z)= \sum_{\boldsymbol{\lambda}}A^{\sigma_1}_{1,\lambda_1}A^{\sigma_2}_{\lambda_1,\lambda_2}A_{\lambda_2\lambda_3,\sigma_3}B^{\sigma_4}_{\lambda_3,\lambda_4}B^{\sigma_5}_{\lambda_4,\lambda_5}B^{\sigma_6}_{\lambda_5,\lambda_6}C^{\sigma_7}_{\lambda_6,\lambda_7}C^{\sigma_8}_{\lambda_7,\lambda_8}C^{\sigma_9}_{\lambda_8,1} \ , \nonumber
\end{equation}
\begin{equation}
   = \sum_{\gamma}\sum_{\boldsymbol{\lambda}}A^{\sigma_1}_{1,\lambda_1}A^{\sigma_2}_{\lambda_1,\lambda_2}U_{\lambda_2\lambda_3,\gamma}S_{\gamma,\gamma}V^{\dag}_{\gamma,\sigma_3}B^{\sigma_4}_{\lambda_3,\lambda_4}B^{\sigma_5}_{\lambda_4,\lambda_5}B^{\sigma_6}_{\lambda_5,\lambda_6}C^{\sigma_7}_{\lambda_6,\lambda_7}C^{\sigma_8}_{\lambda_7,\lambda_8}C^{\sigma_9}_{\lambda_8,1} \ ,
\end{equation}
\end{widetext}
Reshaping $V^{\dag}$ and contracting $U$ and $S$ down into $A^{\sigma_2}$ presents us with Eq.~\eqref{eq:comb_part}, upon relabelling the summation index $\gamma \rightarrow \lambda_2$. The SVD process can then be repeated on the next tensor as in Eq.~\eqref{eq:comb2}.
\begin{widetext}
\begin{equation}
    f(x,y,z)= \sum_{\boldsymbol{\lambda}}A^{\sigma_1}_{1,\lambda_1}M^{\sigma_2}_{\lambda_1,\lambda_2,\lambda_3}\tilde{A}^{\sigma_3}_{\lambda_2}B^{\sigma_4}_{\lambda_3,\lambda_4}B^{\sigma_5}_{\lambda_4,\lambda_5}B^{\sigma_6}_{\lambda_5,\lambda_6}C^{\sigma_7}_{\lambda_6,\lambda_7}C^{\sigma_8}_{\lambda_7,\lambda_8}C^{\sigma_9}_{\lambda_8,1} \ . \label{eq:comb_part}
\end{equation}
\begin{equation}
    f(x,y,z)= \sum_{\boldsymbol{\lambda}}A^{\sigma_1}_{1,\lambda_1}M_{\lambda_1\lambda_3,\sigma_2\lambda_2}\tilde{A}^{\sigma_3}_{\lambda_2}B^{\sigma_4}_{\lambda_3,\lambda_4}B^{\sigma_5}_{\lambda_4,\lambda_5}B^{\sigma_6}_{\lambda_5,\lambda_6}C^{\sigma_7}_{\lambda_6,\lambda_7}C^{\sigma_8}_{\lambda_7,\lambda_8}C^{\sigma_9}_{\lambda_8,1} \ , \label{eq:comb2}
\end{equation}
\begin{equation}
 \implies f(x,y,z)=\sum_{\gamma}\sum_{\boldsymbol{\lambda}}A^{\sigma_1}_{1,\lambda_1}U_{\lambda_1\lambda_3,\gamma}S_{\gamma,\gamma}V^{\dag}_{\gamma,\sigma_2\lambda_2}\tilde{A}^{\sigma_3}_{\lambda_2}B^{\sigma_4}_{\lambda_3,\lambda_4}B^{\sigma_5}_{\lambda_4,\lambda_5}B^{\sigma_6}_{\lambda_5,\lambda_6}C^{\sigma_7}_{\lambda_6,\lambda_7}C^{\sigma_8}_{\lambda_7,\lambda_8}C^{\sigma_9}_{\lambda_8,1} \ , \nonumber
\end{equation}
\begin{equation}
  \implies  f(x,y,z)= \sum_{\boldsymbol{\lambda}}M^{\sigma_1}_{1,\lambda_1,\lambda_3}\tilde{A}^{\sigma_2}_{\lambda_1,\lambda_2}\tilde{A}^{\sigma_3}_{\lambda_3}B^{\sigma_4}_{\lambda_3,\lambda_4}B^{\sigma_5}_{\lambda_4,\lambda_5}B^{\sigma_6}_{\lambda_5,\lambda_6}C^{\sigma_7}_{\lambda_6,\lambda_7}C^{\sigma_8}_{\lambda_7,\lambda_8}C^{\sigma_9}_{\lambda_8,1} \ ,
\end{equation}
\end{widetext}
In doing so we have pushed the $\lambda_3$ bond which connects the $x$ dimension to the $y$ dimension from the final $x$ tensor to the first $x$ tensor, which now forms the comb spine. Repeating this process for the $y$ dimension, and relabelling the $\lambda_3$ and $\lambda_6$ to $\gamma_1$and $\gamma_2$ respectfully will yield us with the following,
\begin{widetext}
\begin{equation}
    f(x,y,z)= \sum_{\boldsymbol{\gamma}}\sum_{\boldsymbol{\lambda}}M^{\sigma_1}_{1,\lambda_1,\gamma_1}\tilde{A}^{\sigma_2}_{\lambda_1,\lambda_2}\tilde{A}^{\sigma_3}_{\lambda_2}M^{\sigma_4}_{\gamma_1,\lambda_4,\gamma_2}\tilde{B}^{\sigma_5}_{\lambda_4,\lambda_5}\tilde{B}^{\sigma_6}_{\lambda_5}M^{\sigma_7}_{\gamma_2,\lambda_7,1}\tilde{C}^{\sigma_8}_{\lambda_7,\lambda_8}\tilde{C}^{\sigma_9}_{\lambda_8} \ ,
\end{equation}
\end{widetext}
which is exactly the connectivity of the comb geometry. Thus we have converted an MPS to a comb geometry via making use of reshaping tensors and SVDs. This process can be understood pictorially as depicted in Fig.~\ref{fig:MPS_to_comb}. This process can be applied for an arbitrary number of dimensions and tensors per dimension. In practice whenever encoding a function into a comb tensor network we take this approach of first encoding it into an MPS with the desired ordering and then exploit SVDs to deform it into a comb tensor network.

\section{Injection boundaries for Electric fields} \label{App:inject}
We wish to be able to inject a laser beam into our plasma domain to study how the propagation of a laser through the plasma can lead to interesting dynamics. In the beat wave simulation of Section \ref{sec:Beat}, we inject two counter propagating plane waves into the plasma, one injected at $x=0$ travelling in the positive $x$ direction, and one at $x=L_x$ travelling in the negative $x$ direction. To be able to implement both lasers, we model each laser propagation independently as its own variable, i.e. a separate field for the laser injected at $x=0$ for that at $x=L_x$. The total laser electric field at any later time is given by the addition of these fields together. 

We outline to procedure to implement such boundaries for the case of the laser injected at the left boundary. Starting from an initial state of $E(t=0)=B(t=0)=0$ at all points, we evolve the fields via Maxwells equations, making use of an appropriate time marching scheme, e.g.,
\begin{equation}
    E(t+dt)= E(t)-dt \frac{\partial B(t)}{\partial x}  \ , \ B(t+dt)= B(t)-dt \frac{\partial E(t)}{\partial x}  \ .
\end{equation}
We have outlined the above for an Euler time-stepping procedure, however we make use of a fourth order Runge-Kutta scheme in our beat wave simulations. All information about the boundary behaviour comes from the spatial derivative with respect to the direction of propagation, $x$. We wish to implement the boundaries such that we inject the laser at $x=0$, but allow the laser to outflow freely at the $x=L_x$ boundary. To do so we utilise a standard second order central finite difference scheme for $\partial_x$ in the bulk, but use a backward finite difference approximation on the final grid-point in the $x$ dimension. This allows the wave to freely propagate outwith the simulation domain once it reaches the opposing end. We can express this form of $\partial_x$ in MPO format as follows,
\begin{widetext}
\begin{eqnarray}
    \hat{\partial}_x 
    =\frac{1}{2\Delta x} \begin{pmatrix}
        s^+_1 && s^-_1 && I_1 &&Y_1
    \end{pmatrix} \begin{pmatrix}
        s^-_2 && 0 && 0 &&0\\
        0&&s^+_2 && 0&& 0\\
        s^+_2 &&s^-_2 && I_2 &&0 \\
        0 && 0 && 0 &&Y_2
    \end{pmatrix} 
    \cdots
     \begin{pmatrix}
        s^-_{N-1}&& 0 &&0&&0\\
        0&&s^+_{N-1} && 0&&0\\
        s^+_{N-1}&&s^-_{N-1} && I_{N-1}&&0\\
        0&&0&&0&&Y_{N-1}
    \end{pmatrix}
    \begin{pmatrix}
        s^-_N \\
        -s^+_N \\
        s^+_N - s^-_N  \\
        2Y_N-s^-_N
        
    \end{pmatrix},
\end{eqnarray}
\end{widetext}
where we have used effective raising and lowering operators at the $i$th site of the MPS,
\begin{equation}
    s_i^{-}=\begin{pmatrix}
        0 & 1 \\
        0 & 0
    \end{pmatrix} \ , \ s_i^{+}=\begin{pmatrix}
        0 & 0 \\
        1 & 0
    \end{pmatrix} , \ I_i=\begin{pmatrix}
        1 & 0 \\
        0 & 1
    \end{pmatrix},Y_i=\begin{pmatrix}
        0 & 0 \\
        0 & 1
    \end{pmatrix}.  
\end{equation}

The action of $\partial_x$ on the first grid point at $x=0$ will be $\hat{\partial_x} E(x=0) = \frac{E(x=\Delta x)}{2\Delta x}$, as there is no grid point below $x=0$. To inject the laser we effectively add an additional grid-point just outside the simulation domain, known as a ghost cell, from where the laser emerges and thus has the time dependent value of $E_o \sin(\omega_L t)$, and means the central finite difference approximation on the $x=0$ grid should take the form of $\frac{ E(x=\Delta x)- E_o\sin(\omega_Lt) }{2\Delta x}$.  Hence the finite difference spatial derivative along the $x$ dimension with an incoming plane wave can be calculated as,
\begin{equation}
    \frac{\partial E}{\partial x} = \hat{\partial_x} E - \frac{E_o \sin(\omega_L t)}{2 \Delta x} \Psi_L(x) \ ,
\end{equation}
where we have defined the state $\Psi_L(x)$ as, 
\begin{align}
\Psi_L=\left\{\begin{array}{ll}
                  1, & x=0,\\[0.1cm]
                 0, & else \ .\\[0.1cm]
                 \end{array}\right. 
\end{align} 
One can apply the same technique for the magnetic fields, and for the right hand side boundary, allowing us to inject EM waves into the plasma simulation. For our beat wave simulations, we desired a laser with a finite extent in the $y$ dimension. To achieve this without introduction of additional complication we chose to inject plane laser waves as outlined above, but then multiply the resultant laser field with a Gaussian envelope along the $y$ dimension before interaction with the plasma to achieve an approximation to a Gaussian wave propagation.

\section{Division of density MPS} \label{app:inverse_den}

\begin{figure}
    \centering
    \includegraphics[width=\linewidth]{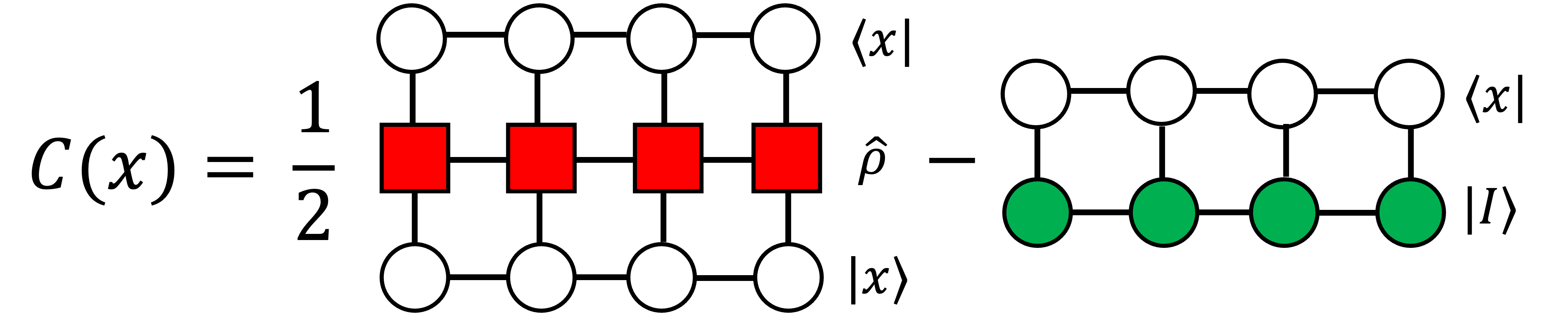}
    \caption{Pictorial representation of the cost function in terms of tensor network states for the inverse density MPS. The current density is encoded as a diagonal MPO shown as red squares, whilst the current estimate for the inverse density is represented by the MPS written with white circles. The MPS consisting of green circles is the identity MPS with bond dimension of unity which evaluates to one at every grid point. We note that for this application each MPS and MPO encodes real values functions and thus $x^*=x$. Furthermore, the MPO $\hat{\rho}$ is diagonal and thus is symmetric and positive definite, as density must always be positive.  }
    \label{fig:DMRG_cost}
\end{figure}

During MHD evolution, we wish to be able to obtain the flow velocity variables $u_i$ in MPS format however the dynamical variable we are time evolving is that of $\rho u_i$, and thus we require the ability to perform a division between two tensor network states. One could contract out the tensor networks into arrays, perform the division and then decompose back into MPS format, but this will be a costly procedure and negates the benefit of using a tensor network approach. Instead  at every time-step, we could solve a variational problem to obtain the inverse density $\frac{1}{\rho}$ as an MPS. We define a cost function $C(x)$ to be minimised as,
\begin{equation}
    C(x)= \frac{1}{2} \bra{x} \hat{\rho }\ket{x} -\bra{x}{I}\rangle   = \frac{1}{2}\sum_{i,j} x_i \hat{\rho}_{i,j}x_j - \sum_j x_j I_j \ , \label{eq:cost}
\end{equation}
where $\ket{x}$ is a real MPS which we are optimising over, $I$ is the Identity MPS, corresponding to the state which is unity at all grid points, and $\hat{\rho}$ is a diagonal MPO with $\rho$ on the diagonal elements (we promote the density MPS to a diagonal MPO). As the diagonal density $\hat{\rho}$ is a symmetric and positive definite operator, the minima of $C(x)$ then occurs for $\hat{\rho}\ket{x}=\ket{I}$ and thus $\ket{x}$ is the solution for the inverse density, i.e.  $x=\frac{1}{\rho}$. The above cost function can more intuitively be viewed in tensor network pictorial form as illustrated in Fig.~\ref{fig:DMRG_cost}.

\begin{figure}
    \centering
    \includegraphics[width=\linewidth]{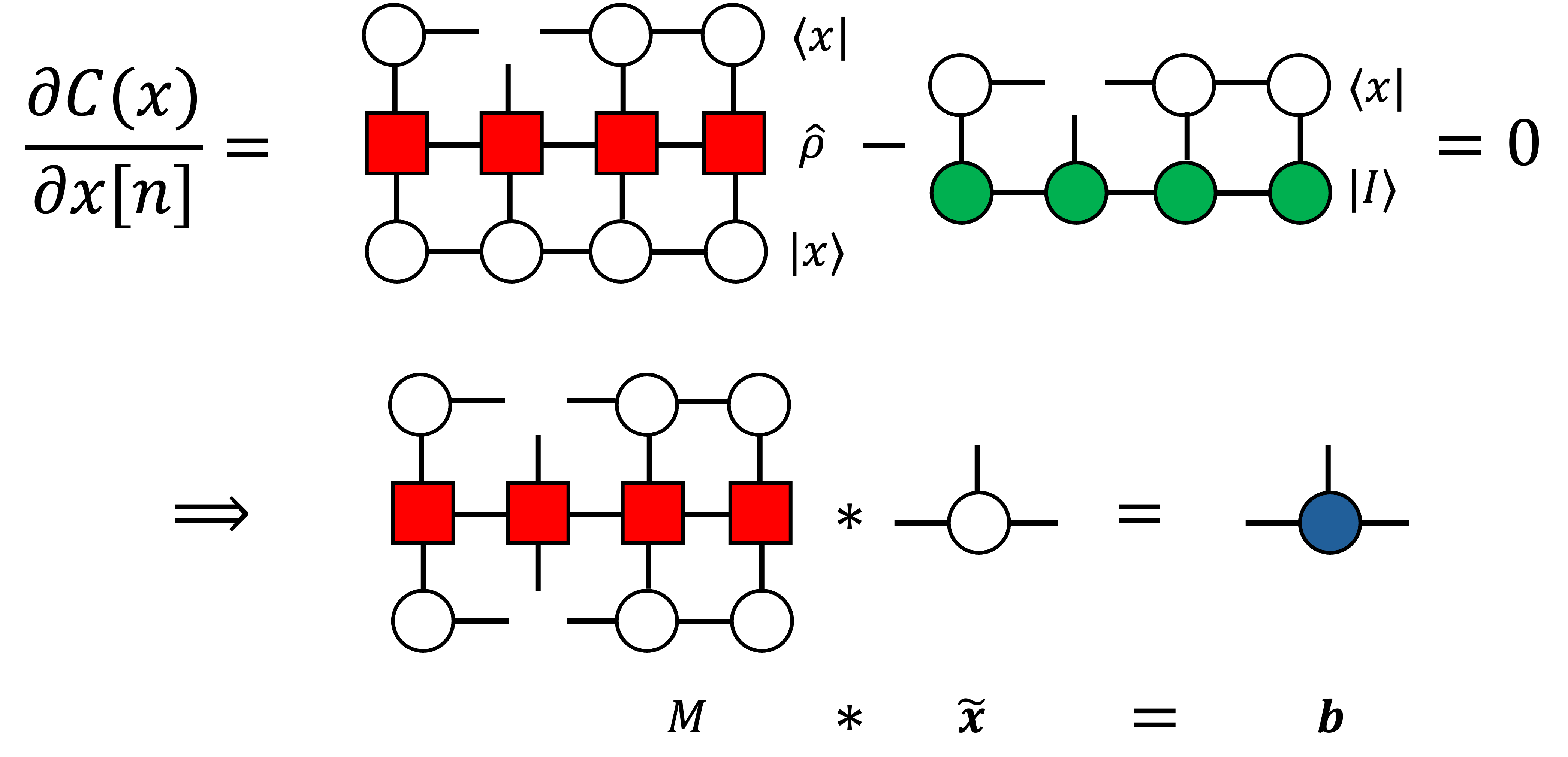}
    \caption{Schematic for minimisation of cost function to obtain the inverse density as an MPS. One optimises each tensor in $x$ individually to obtain a new update $x[n]$ for the optimised tensors at each site. The optimisation reduces to the calculation of a local linear problem in the form of $M\mathbf{\tilde{x}}=\mathbf{b}$, where $M$ is a positive definite matrix, $\mathbf{b}$ is a vector calculated from the current guess MPS, and $\mathbf{x}$ is the update tensor which locally minimises $C(x)$. By sweeping across the MPS many times, one can converge $C(x)$ which then results in $x$ as an MPS approximation to the inverse density.  }
    \label{fig:DMRG_opt}
\end{figure}

One can minimise the above cost function using DMRG like approaches for our MPS \cite{DMRG}, which we outline now. To solve this minimisation, we choose to minimise $C(x)$ with respect to one tensor of $x$ at a time, $x^{\sigma_n}_{\alpha_{n-1},\alpha_n}=x[n]$. We differentiate $C(x)$ with respect to this tensor and set the derivative to zero,
\begin{equation}
    \frac{\partial C(x)}{\partial x[n]}=0 \ . \label{eq:opt}
\end{equation}
The optimisation scheme is most easily outlined in tensor network pictorial form, shown in Fig.~\ref{fig:DMRG_opt}, where we can see that Eq.~\eqref{eq:opt} reduces to solving a straightforward linear problem of the form $M \mathbf{\tilde{x}}=\mathbf{b}$, where $M$ is a positive semi-definite matrix. One can solve this local linear equation via standard Krylov based approaches. Having obtained the new optimised tensor (which corresponds to $\mathbf{\tilde{x}}$ in the local linear problem), this is then substituted into the corresponding site in the MPS. Having optimised one tensor in the MPS, this process is then repeated for the remaining tensors in order to update our solution for the inverse density. 

In practice, one begins at either the leftmost or rightmost tensor in the MPS, performs the optimisation, and then shifts focus to the next tensor along the chain, where the next optimisation is performed and so on until one reaches the opposite end of the MPS. Typically one would then perform the same process in the reverse order to constitute one full sweep of the DMRG like optimisation. One then performs multiple of these sweeps until sufficient convergence criteria of the cost function is obtained such that the solution to $x$ is the inverse density in MPS format. The outlined scheme above is known as a one-site DMRG scheme, where the bond dimension of $x$ is controlled by the initial guess MPS and does not change during the optimisation. Often a two-site scheme is employed in which two tensors at a time are optimised. This has the advantage of being able to adaptively control the bond dimension during optimisation, and can often help reduce the likely hood of being stuck in a local minima \cite{DMRG}. One can control the desired level of accuracy for the inverse density via modifying the bond dimension and the tolerance of the local Krylov solver.

\section{Shuman filters for MHD} \label{App:Shuman}

When performing simulations of MHD whilst using an RK4 scheme, we additionally make use of a form of digital filter known as a Shuman filter \cite{Shapiro}. Such a filter is designed to be applied onto data to remove, or at least suppress, small numerical fluctuations which could otherwise lead to instabilities in the simulations. The Shuman filter is defined for a 2D data set $u_{i,j}$ , where $i$ and $j$ index the $x$ and $y$ dimensions receptively, as follows,
\begin{equation}
    \tilde{u}_{i,j}=\frac{u_{i-1,j} + u_{i,j-1} +\beta u_{i,j} + u_{i,j+1} +u_{i+1,j}   }{4+\beta} \ .
\end{equation}
The parameter $\beta$ is used to control the extent of filtering applied. In the limit of $\beta\rightarrow\infty$, the Shuman filter collapses to $\tilde{u}_{i,j}=u_{i,j}$ i.e. no filtering is applied. Instead for $\beta=0$, each data point is replaced by the average of its four nearest neighbours.

One can write the Shuman filter as an MPO $\hat{F}_{\beta}$ to act on an MPS encoding data as follows,
\begin{equation}
   \hat{F}_{\beta}= \frac{1}{4+\beta} \left[\beta \hat{I} +\hat{S}^{+}_x +\hat{S}^+_y +\hat{S}^-_x +\hat{S}^-_y   \right] \ ,
\end{equation}
where $I$ is the identity operator, $S^{+}_l$ is the raising MPO, shifting all points to the right, and $S^{-}_l$ the lowering MPO. Applying the above MPO onto each MPS in our plasma simulation after each time-increment allows us to filter the data to improve the stability and smooth out sharp features present. 
\begin{figure*}[t]
    \centering
    \includegraphics[width=0.8\linewidth]{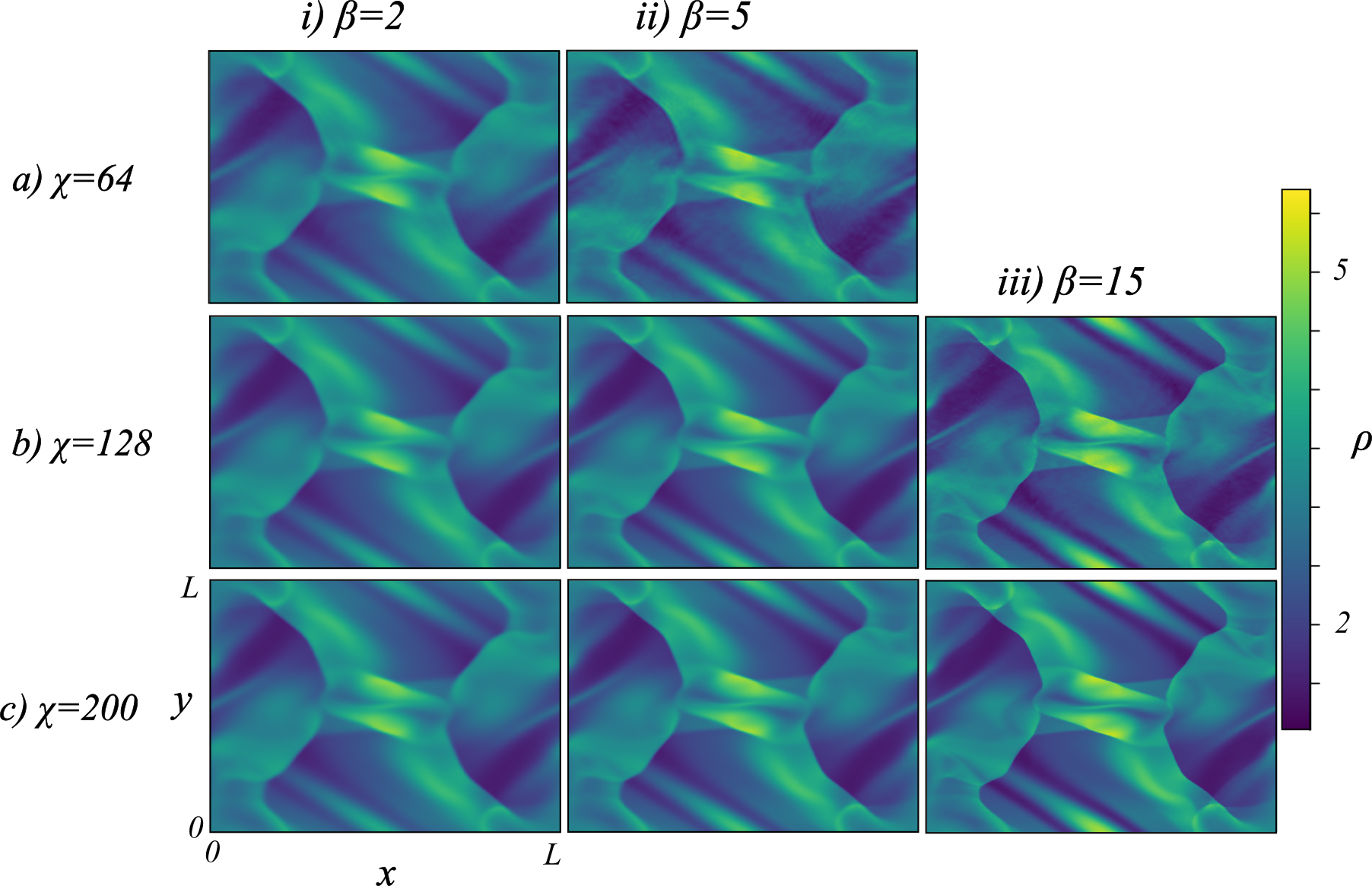}
    \caption{Snapshots of the Orszag-Tang vortex on a $2^9\times2^9$ spatial grid at a simulation time of $t=3$, calculated via making use of different filtering parameters with fixed bond dimensions of $a)$ $\chi=64$,$b)$ $\chi=128$ and $\chi=200$. After each time-increment of the RK4 evolution scheme the filtering MPO is applied, resulting in solutions of varying degrees of smoothness dependant on the parameter $\beta$, with smaller $\beta$ yielding smoother solutions. Note that $\chi=64, \beta=15$ became unstable and broke down before reaching the desired simulation time, however increasing the bond dimension to $\chi=128$ and beyond allows the dynamics to be recovered. 
    }
    \label{fig:MHD_beta}
\end{figure*}

The need to specify $\beta$ for the Shuman filter introduces an additional control parameter. We analysed the impact of $\beta$ on the dynamics of the Orszag-Tang vortex in Fig.~\ref{fig:MHD_beta} as a function of bond dimension. We largely observe that a smaller $\beta$ tends to result in a much smoother solution, where one can still observe the formation of shock waves, however they do not appear as sharp as when one uses a larger $\beta$. The benefit of a small $\beta$ however is that it allows one to use a smaller bond dimension and still accurately capture the dynamics. We see that a bond dimension of $\chi=64$ almost perfectly matches the solution using $\chi=200$ for $\beta=2$. However as $\beta$ is increased to $\beta=5$, we begin to see a large amount of noise present in the $\chi=64$ simulation. Increasing to $\beta=15$ further for $\chi=64$ allowed the simulation to run for some time however the simulation broke down before it could be run for long enough to observe the dynamics at $t=3$. At $\beta=15$, a bond dimension of $\chi=128$ does manage to run fully, still capturing the relevant shock formation, but one does still notice a small amount of noise. Only at $\chi=200$ for $\beta=15$ does one achieve a clean solution. This indicates that the apparent smoothness of the solution has a large impact on the required bond dimension to capture the dynamics, with smoother solutions requiring fewer numerical resources to accurately capture.

\end{document}